\documentclass[11pt,a4paper,english]{article}
\pdfoutput=1
\usepackage[utf8]{inputenc}
\usepackage[left=0.8in,right=0.8in,top=1.0in,bottom=1.2in]{geometry}
\usepackage[american]{babel}
\usepackage{cite}
\usepackage{amsmath}
\usepackage{amsfonts}
\usepackage{amssymb}
\usepackage{graphicx}
\usepackage{tikz}
\usepackage{pdflscape}				
\usepackage{multirow}				
\usepackage{float}					
\usepackage{hyperref}
\hypersetup{
	unicode=false,					
	pdftoolbar=true,				
	pdfmenubar=true,				
	pdffitwindow=false,				
	pdfstartview={FitH},				
	pdftitle={Notes},				
	pdfauthor={Felix Wilsch},		
	pdfsubject={Subject},			
	pdfcreator={Creator},   		  	
	pdfproducer={pdflatex}, 			
	pdfkeywords={SMEFT, Flavour, U(2)},		 	
	pdfnewwindow=true 				
}
\usepackage{color,amsmath,amssymb,exscale,psfrag,epsfig,colortbl,bm,slashed,array,bbold}
\usepackage{fancyhdr}

\usepackage{graphics}

\definecolor{pink}{rgb}{0.94, 0.5, 0.5}
\definecolor{lightblue}{rgb}{0.39, 0.58, 0.93} 

\newcommand{\be}{\begin{equation}}
\newcommand{\ee}{\end{equation}}
\newcommand{\bea}{\begin{eqnarray}}
\newcommand{\eea}{\end{eqnarray}}
\newcommand{\ba}{\begin{array}}
\newcommand{\ea}{\end{array}}
\newcommand{\no}{\nonumber}
\newcommand{\cL}{\mathcal{L}}
\newcommand{\cO}{\mathcal{O}}
\newcommand{\Vql}{V_{q(\ell)}}
\newcommand{\eps}{\epsilon}
\newcommand{\epsql}{\epsilon_{q(\ell)}}
\newcommand{\epsq}{\epsilon_q}
\newcommand{\epsl}{\epsilon_\ell}
\newcommand{\del}{\delta}
\newcommand{\delp}{\delta^\prime}
\newcommand{\delpsq}{\delta^{\prime 2}}
\newcommand{\SMEFT}{SMEFT}
\newcommand{\gsim}{\lower.7ex\hbox{$\;\stackrel{\textstyle>}{\sim}\;$}}
\newcommand{\lsim}{\lower.7ex\hbox{$\;\stackrel{\textstyle<}{\sim}\;$}}

\begin{document}

\begin{flushright}
\end{flushright}

\begin{center}
\vspace{1.5cm}
     {\Large\bf Flavour symmetries in the SMEFT}
       \\ [1cm] 
   {\bf  Darius A. Faroughy$^{(a)}$, Gino Isidori$^{(a)}$, Felix Wilsch$^{(a)}$, Kei Yamamoto$^{(a,b)}$}    \\[0.5cm]
  {\em $(a)$Physik-Institut, Universit\"at Z\"urich, CH-8057 Z\"urich, Switzerland}  \\
    {\em $(b)$Graduate School of Science, Hiroshima University, Higashi-Hiroshima 739-8526, Japan}
\end{center}
\vspace{1cm}

\begin{quote}
We analyse how $U(3)^5$ and $U(2)^5$ flavour symmetries act on the Standard Model Effective Field Theory, providing an 
organising principle to classify the large number of dimension-six operators involving fermion fields. 
A detailed counting of such operators, at different order in the breaking terms of both these symmetries, is presented.
A brief discussion about possible deviations from these two reference cases, and a simple example of the 
usefulness of this classification scheme for high-$p_T$ analyses at the LHC, are also presented.
\end{quote}

\tableofcontents	
\newpage

\section{Introduction}
\label{sec:intro}

Since its conception in the early 70s, the Standard Model (SM) has been regarded as the low-energy limit of an extended theory that includes more degrees of freedom, 
around or above the electroweak scale, addressing some of its open issues.  
After the first years of running of the LHC we can state with confidence that there is a mass gap between the SM spectrum and these hypothetical, but still highly motivated, additional degrees of freedom. How large is this mass gap, is probably the most interesting open question nowadays in high-energy physics.
  
The observation of a mass gap above the SM spectrum, and the need to describe in general terms possible physics beyond the SM, has motivated the systematic study of what  goes under the name of SMEFT: the Effective Field Theory (EFT) based on the 
$SU(3)_c\times SU(2)_L \times U(1)_Y$ local symmetry, and the SM field content (including the Higgs) 
as dynamical degrees of freedom below a cut-off scale $\Lambda  >  G^{-1/2}_F$.
Several years after the pioneering analysis in~\cite{Buchmuller:1985jz},  the first complete non-redundant classification of baryon- and lepton-number conserving dimension-six operators in the SMEFT has been presented in~\cite{Grzadkowski:2010es}. Employing such basis, the Renormalization Group (RG) evolution of the Wilson coefficients of  these operators, at the one-loop level, has been analysed in~\cite{Jenkins:2013zja,Jenkins:2013wua,Alonso:2013hga}\footnote{See  
e.g.~\cite{Brivio:2017vri,  Falkowski:2017pss, Descotes-Genon:2018foz,	Falkowski:2019hvp,  Aoude:2020dwv} for recent reviews and phenomenological analyses 
of the SMEFT with special emphasis on flavour observables.}

While the number of independent electroweak structures amounts to less than one hundred terms, a large proliferation in the number of independent terms (and corresponding  coefficients) in the SMEFT occurs when all the possible flavour structures are taken into account: in absence of any flavour symmetry, they amount to  1350 CP-even and 1149 CP-odd independent coefficients for the dimension-six operators~\cite{Alonso:2013hga}.  
The purpose of the present paper is to analyse how a series of motivated hypothesis about flavour symmetries and symmetry-breaking terms can help reduce, and order via an appropriate power counting, such large number of independent terms.

The need of specific hypothesis about symmetry and symmetry-breaking in the flavour sector plays an important role in addressing the key question of how large is the mass gap above the SM spectrum,  or the cut-off scale of the SMEFT.  In a na\"\i ve flavour-anarchic approach, the bounds on the dimension-six operators are dominated by those contributing at tree-level to flavour-violating observables, in particular to $\Delta F=2$ and lepton-flavour violating processes. These operators set bounds of $\cO(10^{5})$~TeV on $\Lambda$ for  $\cO(1)$ coefficients~\cite{Isidori:2010kg}. If this high scale were the overall cut-off scale of the SMEFT, it would imply a severe fine-tuning problem on the Higgs mass term, and it would also imply that most of the operators in the SMEFT play an irrelevant role in current experiments (making the whole construction impractical from the phenomenological point of view). On the other hand,  from the known structure of the SM Yukawa couplings, we know that flavour is highly non generic, 
at least in the dimension-four sector of the EFT. It is therefore natural to employ specific hypothesis about flavour symmetry and symmetry-breaking terms on the whole SMEFT. This procedure has multiple advantages:  
i)~it allows us to lower the overall cut-off scale of the EFT, ameliorating the fine-tuning problem on the Higgs mass; 
ii)~it reduces the number of independent parameters;~iii) 
it makes the EFT construction more consistent and somehow more  ``appealing", with competing constraints from  
flavour-conserving and flavour-violating processes on a given effective operator. 

The price to pay for this series of advantages is the choice of the flavour symmetry (and symmetry-breaking sector), which necessarily introduces some model dependence. 
However, if we are interested in symmetries and symmetry-breaking patterns able to successfully reproduce the SM Yukawa couplings and, at the same time, suppress 
non-standard contributions to flavour-violating observables, the choice is limited.  In this paper we focus on two main cases which are particularly 
motivated from this point of view, i.e.~the flavour symmetries $U(3)^5$ and $U(2)^5$, with possible minor variations. 
The $U(3)^5$ flavour symmetry is the maximal flavour symmetry allowed by the SM gauge group, while $U(2)^5$ is the corresponding 
subgroup acting only on the first two (light) generations. The $U(3)^5$ symmetry allows us to implement the Minimal Flavour Violation (MFV)
hypotheses~\cite{Chivukula:1987py,DAmbrosio:2002vsn}, which is the most restrictive consistent hypothesis we can utilize in the SMEFT 
to suppress non-standard contributions to flavour-violating observables~\cite{DAmbrosio:2002vsn}. The $U(2)^5$ symmetry with minimal breaking~\cite{Barbieri:2011ci,Barbieri:2012uh,Blankenburg:2012nx}
is quite interesting since it retains most of the MFV virtues, but it allows us to have a much richer structure as far as third-generation dynamics is 
concerned. For instance, as pointed out in~\cite{Greljo:2015mma,Barbieri:2015yvd,Buttazzo:2017ixm}, 
the $U(2)^5$ setup  provides a very efficient EFT description of the recent flavour anomalies, 
which cannot be accommodated within a MFV framework. 

It must be stressed that the flavour symmetries we are considering are not necessarily fundamental symmetries of the ultraviolet (UV) theory. 
They could well be accidental symmetries, associated e.g.~to some underlying dynamics that act in a non-universal way on the different generations. From the EFT point of view,  we cannot distinguish between  fundamental or dynamical symmetries in the UV: both of them are effectively described by imposing a specific (global) flavour symmetry in the EFT and specifying a well-defined set of symmetry-breaking terms (the spurions). This effective  
approach is the one we employ in the present analysis. 

It is clear that the $U(3)^5$ and $U(2)^5$ symmetries are not the only options to  efficiently suppress flavour-violating observables in the SMEFT.
An interesting alternative is provided by the ample class of models based on $U(1)$ symmetries \`a la Froggatt-Nielsen~\cite{Froggatt:1978nt}.\footnote{An interesting 
systematic analysis about the implementation of $U(1)$ symmetries in the SMEFT, together with general dynamical assumptions about
new physics, has recently been presented in Ref.~\cite{Bordone:2019uzc}.} However, on the one hand non-Abelian symmetries 
are more predictive in establishing a link between SM Yukawa couplings and flavour-violating effects in the SMEFT.
On the other hand, in the absence of suprions of non-Abelian groups, a classification of operators such as the one presented 
here is not particularly illuminating. This is why in this paper we restrict the attention to the case of $U(3)^5$ and $U(2)^5$
symmetries with minimal breaking.

The paper is organised as follows. In Section~\ref{sec:U3} we analyse the $U(3)^5$ symmetry and the MFV hypothesis.
In Section~\ref{sec:U2} we  analyse the $U(2)^5$ symmetry with minimal breaking.
A general discussion about possible deviations from these two reference cases, with the specific analysis 
of the impact of $U(1)$ symmetries acting on third-generation down-quarks and/or charged leptons,
is presented in Section~\ref{sect:beyondU2}. In Section~\ref{sect:pheno} we briefly illustrate the usefulness of this 
approach in high-$p_T$ phenomenological analyses at the LHC, choosing $pp\to \ell \tau$ ($\ell=e,\mu$) as a 
representative example.
The results are summarised in the Conclusions.

\bigskip

\section{The $U(3)^5$ symmetry and MFV}
\label{sec:U3}
The largest group of flavour-symmetry transformations compatible with 
the gauge symmetries of the SM Lagrangian is~\cite{Chivukula:1987py}
\be
	G_{\rm flavour} = U(3)^5 = U(3)_\ell \otimes U(3)_q \otimes U(3)_e \otimes U(3)_u \otimes U(3)_d 
	= SU(3)^5 \otimes U(1)^5~,
\ee
where, with a standard notation, $\{\ell, q, e, u, d\}$ denote the five independent types of SM fermions 
with different gauge quantum numbers:
\be
\cL^{\rm fermions}_{\rm SM} =\sum_{\psi = \ell, q, e, u, d} \bar \psi i \slash{\!\!\!\! D} \psi + \left( \bar \ell Y_e e H +  \bar q Y_d d H +  \bar q Y_u u H_c + 
~{\rm h.c.} \right). 
\label{eq:SMfermions}
\ee
Within the SM, the Yukawa couplings ($Y_{e,u,d}$) are the only 
source of breaking of $G_{\rm flavour}$.  They break this global symmetry as follows 
\bea
G_{\rm flavour} = \left\{
\ba{l}
SU(3)^5 \\[3pt]   U(1)^5 
\ea \right.
\stackrel{ Y_{e,u,d}\neq0 }{\longrightarrow} 
\ba{l}
 U(1)_{e - \mu} \otimes U(1)_{\tau - \mu}  \\
 U(1)_B \otimes U(1)_L \otimes U(1)_Y  \\
\ea
\eea
where we separated explicitly flavour-universal  and flavour-non-universal subgroups.
The three unbroken flavour-universal $U(1)$ groups are baryon number, lepton number, and 
hypercharge.\footnote{The two flavour-non-universal $U(1)$ subgroups
left unbroken by the Yukawa couplings are arbitrary combinations of the two diagonal 
generators of $SU(3)_{e+\ell}$, namely the vectorial subgroup of  $SU(3)_e \times SU(3)_\ell$.}

As anticipated, our goal is to count  and classify the number of independent  dimension-six operators, 
and corresponding effective couplings,
in the \SMEFT\ according to different hypotheses about the breaking of $G_{\rm flavour}$. 
The most restrictive assumption we can make is that $G_{\rm flavour}$ is an exact symmetry 
of the beyond-the-SM sector. This assumption is not fully consistent, since $G_{\rm flavour}$
is broken within the SM. However, it is a useful starting point for the classification of the operators,
and it is a coherent hypothesis to be implemented in the SMEFT in the limit where we neglect $G_{\rm flavour}$ breaking terms also in the SM sector,
i.e.~in the limit where we neglect the SM Yukawa couplings.

In this section we compare the results obtained in this limit (i.e.~the exact $U(3)^5$ limit), as well as 
those obtained under the MFV hypothesis (considering the first few terms in the expansion in powers of Yukawa couplings),
to those obtained in the absence of any flavour symmetry assuming one or three generations
of SM fermions.

\paragraph{Exact $U(3)^5$ symmetry.}
To classify the \SMEFT\ operators we adopt the Warsaw basis~\cite{Grzadkowski:2010es}, whose notation to identify 
the different electroweak structures, adopting the division in classes introduced in~\cite{Alonso:2013hga},
will be followed throughout the whole paper.\footnote{To facilitate the readability 
of this paper, the complete list of operators containing  fermion fields is reported in Table~\ref{tab:OperatorClasses} of Appendix~\ref{app:tables}.}
The operators of classes~1-4 do not contain fermions, thus the counting of independent couplings is trivial: 9 independent CP-even 
coefficients for the 9 hermitian structures and 6 CP-odd coefficients for the  anti-hermitian ones.
The operators of the classes 5 and 6 are forbidden in the exact $U(3)^5$ limit, since they contain a fermion 
current of the type $\bar{L}R$. The class 7 operators, but for $Q_{Hud}$, are all hermitian and allowed, 
provided the fermion indices are properly summed, hence they contribute one real coefficient each. 
The operator $Q_{Hud}$ is forbidden.

For the operators containing four fermions we have three structures which are hermitian,
namely $(\bar{L}L)(\bar{L}L)$, $(\bar{R}R)(\bar{R}R)$ and $(\bar{L}L)(\bar{R}R)$,
which are allowed by the symmetry provided the fermion indices are properly summed.
Here each operator corresponds to one real coefficient, with the exception of 
$Q_{\ell\ell}$, $Q_{qq}^{(1)}$, $Q_{qq}^{(3)}$, $Q_{uu}$ and $Q_{dd}$, which corresponds to two independent
operators (hence two real coefficients) since there are two independent $U(3)^5$-invariant 
ways to contract the flavour indices. For example, in the case of $Q_{\ell\ell}$ we find 
\begin{align}
	\left( \bar{\ell}_p \gamma_\mu \ell_p \right) \left( \bar{\ell}_r \gamma_\mu \ell_r \right) && \text{and} && \left( \bar{\ell}_p \gamma_\mu \ell_r \right)\left( \bar{\ell}_r \gamma_\mu \ell_p \right)~,
\end{align}
where $r$ and $p$ denote the flavour indices (and the sum over repeated indices is understood).
Note that the $Q_{ee}$ operator only corresponds to a single independent structure due to the Fierz identity
\begin{align}
	\left( \bar{e}_p \gamma_\mu e_r \right) \left( \bar{e}_s \gamma_\mu e_t \right) = \left( \bar{e}_s \gamma_\mu e_r \right) \left( \bar{e}_p \gamma_\mu e_t \right).
	\label{eq:Fierz}
\end{align}
The operators of the type $(\bar{L}R)(\bar{R}L)$ and $(\bar{L}R)(\bar{L}R)$ are not allowed by the symmetry.

The results thus obtained are reported in Table~\ref{tab:U3new} in the ``Exact" $U(3)^5$ column: the left (right) value in each entry indicates 
the number of CP-even (CP-odd) coefficients. For comparison, we also show the counting of independent
coefficients  if no symmetry is imposed, or if a single generation of fermions is considered, where we fully agree with 
the results derived first in Ref.~\cite{Alonso:2013hga}. 
The counting in the latter case proceeds in close analogy to the $U(3)^5$ case with a few important 
differences: the operators in classes 5 and 6, as well as $Q_{Hud}$ in class 7,
which are not hermitian, leads to one real and one imaginary coefficients each. 
The operators  in the   $(\bar{L}L)(\bar{L}L)$ and $(\bar{R}R)(\bar{R}R)$ categories  which had two possible 
flavour contractions with more generations, now have only a single contraction, hence in these 
categories we have one real coefficient for each electroweak structure.  
Finally, for the non-hermitian operators of the type $(\bar{L}R)(\bar{R}L)$ and $(\bar{L}R)(\bar{L}R)$ 
we can identify one real and one imaginary coefficient for each electroweak structure.

\begin{table}[t]
\begin{center}
	\renewcommand{\arraystretch}{1.2} 
	\begin{tabular}{c   l   ||   l   l   |   l   l   ||      l   l   |   l   l  |    l  l    }
	& &   \multicolumn{4}{c||}{	No symmetry }  &  \multicolumn{6}{c}{ $U(3)^5$}   \\
   Class & Operators & \multicolumn{2}{c|}{  3 Gen.}  & \multicolumn{2}{c||}{  1 Gen.}  &  \multicolumn{2}{c|}{  Exact } &  \multicolumn{2}{c|}{ $\cO(Y_{e,d,u}^1)$ }   &  \multicolumn{2}{c}{ $\cO(Y_{e}^1, Y_d^1 Y^2_u)$ }  \\  \hline
		1--4 			& $X^3$, 	$H^6$, $H^4 D^2$, $X^2 H^2$		& 9 		& 6 		& 9	& 6 	& 9	& 6  	& 9	& 6   &  9  &  6   \\	\hline
		5 					& $\psi^2 H^3$ 			& 27 		& 27 		& 3	& 3	& -- 	& -- 	& 3	& 3	&  4	& 4	\\	\hline	
		6 					& $\psi^2 X H$ 		& 72		& 72		& 8	& 8	& --	& --	& 8	& 8	&  11	& 11	\\
		7 					& $\psi^2 H^2 D$			 	& 51		& 30		& 8	& 1	& 7	& --	& 7	& --    & 11	& 1  \\ \hline
		\multirow{5}{*}{8}		& $(\bar{L}L)(\bar{L}L)$		 	& 171	& 126	& 5	& --	& 8	& --    & 8	& --	& 14 &	-- \\
							& $(\bar{R}R)(\bar{R}R)$		 	& 255	& 195	& 7	& --	& 9	& --	& 9	& --	& 14 &	-- \\
							& $(\bar{L}L)(\bar{R}R)$		 	& 360	& 288	& 8	& --	& 8	& --	& 8	& --	& 18 &	-- \\
							& $(\bar{L}R)(\bar{R}L)$ 		& 81		& 81		& 1	& 1	& --	& --	& --	& --	& --	& --	\\	
							& $(\bar{L}R)(\bar{L}R)$ 	 	& 324	& 324	& 4	& 4	& --	& --	& --	& --	& 4	& 4	\\ \hline
		\multicolumn{2}{c||}{\bf total:}					 		&1350	& 1149	& 53	& 23 & 41	& 6	& 52 & 17	& 85 & 26
			\end{tabular}
	\caption{Number of independent operators in $U(3)^5$, MFV and without symmetry. In each column the left (right) number corresponds to the number of CP-even (CP-odd) coefficients.  
	$\cO(X^n)$ stands for including  terms up to $\cO(X^n)$. 
	\label{tab:U3new}}
	\end{center}
\end{table}

\subsection{Minimal Flavour Violation.}
The MFV hypothesis is the assumption that the SM Yukawa couplings are the only sources of $U(3)^5$ breaking~\cite{Chivukula:1987py,DAmbrosio:2002vsn}. 
The exact $U(3)^5$ limit analysed before is equivalent to employing the MFV hypothesis and working to zeroth order in the symmetry breaking terms.
To go beyond leading order we promote the SM Yukawa couplings to $U(3)^5$ spurion fields with the following 
transformation properties~\cite{DAmbrosio:2002vsn}:
\begin{align}
	Y_u=\left( 1,3,1, \bar{3},1 \right)~, && 	Y_d=\left( 1,3,1,1,\bar{3} \right)~, &&	Y_e=\left( 3,1,\bar{3},1,1 \right)~.
\end{align} 
In principle, the spurions can appear with arbitrary powers both in the renormalizable ($d=4$) part of the Lagrangian and in the dimension-six effective operators. 
However, via a suitable redefinition of both fermion fields and spurions, we can always put the $d=4$ Lagrangian to its standard expression
in Eq.~(\ref{eq:SMfermions}), namely we can always identify the spurions with the SM Yukawa couplings. 
This implies we can always choose a flavour basis where the spurions are 
completely determined in terms of fermion masses and the Cabibbo-Kobayashi-Maskawa (CKM) matrix, $V_{\rm CKM}$. A representative 
example is the down-quark mass-eigenstate basis, where
\be
Y_e =  \textrm{diag}(y_e,y_\mu,y_\tau)~, \qquad
Y_d  = \textrm{diag}(y_d,y_s,y_b)~, \qquad
Y_u =  V_{\rm CKM}^\dagger \times \textrm{diag}(y_u,y_c,y_t)~.
\label{eq:d-basis}
\ee
The key point is that there are no free (observable) parameters in the structure of the MFV spurions. As we shall see, this is not the case 
for less restrictive symmetry hypotheses, such as the $U(2)^5$ case discussed in sect.~\ref{sec:U2}.
We are now ready to count the number of independent operators appearing at $d=6$ in the SMEFT
inserting a small number of symmetry breaking terms. 

\paragraph{Terms of $\mathcal{O}(Y_{u,d,e})$.}
With a single insertions of the Yukawa couplings, 
only the operators in class 5 and 6 gets modified 
with respect to the $U(3)^3$ invariant case: as far as the flavour structure is concerned, these operators are identical  
to the three Yukawa interactions in Eq.~(\ref{eq:SMfermions}). Since they are not hermitian, we get 3~(8)~CP-even and 3~(8)~CP-odd parameters 
for $\psi^2 H^3$+h.c. ($\psi^2 X H$+h.c.). The counting of independent terms thus obtained, 
reported in Table~\ref{tab:U3new}, is consistent with that performed in~\cite{Brivio:2017btx}.

\paragraph{Terms of $\mathcal{O}(Y_u^2)$.} Here the operators involved are those in class 7 and 8, which contain at least
two $q$ fields or two $u$ fields, and that we can conveniently re-arrange in the following three categories
\begin{itemize}
\item[A)] Operators with a bilinear current of the type $\bar q \Gamma q$ or $\bar u  \Gamma u$:
\begin{description}
\item{} 3 in class~7: $Q_{Hq}^{(1,3)}$	and $Q_{Hu}$  
\item{} 2 in class~$(\bar{L}L)(\bar{L}L)$:   $Q_{\ell q}^{(1,3)}$   
\item{} 3 in class~$(\bar{R}R)(\bar{R}R)$:   $Q_{eu}$ and $Q_{ud}^{(1,8)}$  
\item{} 4 in class~$(\bar{L}L)(\bar{R}R)$:  $Q_{\ell u}$,  $Q_{qe}$, $Q_{qd}^{(1,8)}$ 
\end{description}
\item[B)] Operators of the type $\bar q \Gamma q \times \bar u  \Gamma u$:
\begin{description}
\item{} 2 in class~$(\bar{L}L)(\bar{R}R)$:	$Q_{qu}^{(1,8)}$ 
\end{description}
\item[C)] Operators with four $u$ or four $q$ fields:
\begin{description}
\item{} 2 in class~$(\bar{L}L)(\bar{L}L)$:  $Q_{qq}^{(1,3)} $
\item{} 1 in class~$(\bar{R}R)(\bar{R}R)$:  $Q_{uu}$
\end{description}
\end{itemize}
where $\Gamma$ denote a generic combination of Dirac matrices, color and $SU(2)_L$ generators, which play no role as far as the 
flavour structure is concerned. 
For the operators in the category A) we obtain a $U(3)^5$ singlet contracting $Y_u$ and $Y_u^\dagger$ 
to form an octet of $SU(3)_u$  or  $SU(3)_q$, and then contracting this octet with the flavour indices of the 
$q$- or $u$-quark current (the other current being necessarily a flavour singlet):
\be
 \bar{q}_p \Gamma q_r   ( Y_u Y_u^\dagger)_{pr}~,
\qquad 
\bar{u}_p \Gamma  u_r  (Y_u^\dagger Y_u)_{pr}~.
\ee
Thus all the hermitian structures in the category A) yield one CP-even coupling.
For the operators in the category B) three contractions are possible:
\be
 (\bar{u}_a \Gamma u_a )  (\bar{q}_p \Gamma q_r )  ( Y_u Y_u^\dagger)_{pr}~, \qquad 
  (\bar{u}_p \Gamma u_r )  (\bar{q}_a \Gamma q_a )  (Y_u^\dagger Y_u)_{pr}~, \qquad 
   (\bar{u}_p \Gamma u_s )  (\bar{q}_r \Gamma q_t)  (Y_u^\dagger)_{pt}  (Y_u)_{rs} ~,
\ee
Thus all the hermitian structures in the category B) yield three CP-even couplings.
Finally, for the operators in the category C) only two contractions are possible, such as 
\be
 (\bar{q}_a \Gamma q_a )  (\bar{q}_p \Gamma q_r )  ( Y_u Y_u^\dagger)_{pr}~, \qquad 
   (\bar{q}_p \Gamma q_a )  (\bar{q}_a \Gamma q_r) ( Y_u Y_u^\dagger)_{pr}~, \qquad  
\ee
and similarly for the $u$ fields.  All other contractions either reduce to those or to genuine singlet
contractions that have already been counted in the $U(3)^5$ invariant case. Thus 
all the hermitian structures in the category C) yield two CP-even couplings.
Summing up, we find the following CP-even couplings for operators 
with two powers of $Y_u$: 3  in class~7, 6  in class~$(\bar{L}L)(\bar{L}L)$,
5  in class~$(\bar{R}R)(\bar{R}R)$ and 10 in class~$(\bar{L}L)(\bar{R}R)$.

\paragraph{Terms of $\mathcal{O}(Y_u Y_d)$ and $\mathcal{O}(Y^2_u Y_d)$.}
Proceeding in a similar manner we can identify the independent terms with one $Y_d$ and one or two $Y_u$ spurions.
Three non-hermitian structures can have flavour-singlet contractions with one $Y_d$ and one $Y_u$:
$Q_{Hud}$ (1 possibility), $Q_{quqd}^{(1)}$ and $Q_{quqd}^{(8)}$ (two possibilities each), for a total of 
 five CP-even and five CP-odd parameters. Inserting two up-type and one down-type spurions, 
 we can form four (non-hermitian) flavour-singlet operators 
 using the structures with one $d$ and one $q$ fields  in class 5 and 6, 
 for a total of four CP-even and four CP-odd parameters.

\subsection{Summary and discussion}

The overall number of independent terms allowed by the MFV hypothesis with at most one ``small'' Yukawa coupling, namely 
$Y_d$ and $Y_e$, and up to two powers of $Y_u$ is shown in the last column of Table~\ref{tab:U3new}.\footnote{A detailed counting 
order by order in the insertions of different powers of the Yukawa couplings in presented in Table~\ref{tab:MFVdetails}
in Appendix~\ref{app:tables}.}
At this order we have all the operators necessary to describe deviations from the SM in rare flavour-violating 
processes that do occur within the SM and, within the SM, receive sizeable short-distance contributions induced
by the large top-quark mass (such as $B^0$--$\bar B^0$ and $K^0$--$\bar K^0$ mixing, $b\to s \gamma$, $b\to s \ell^+\ell^-$, \ldots) \cite{DAmbrosio:2002vsn}. 
As can be seen, the number of operators at this order is much smaller than that obtained 
in absence of any symmetry (for three generations) and still remarkably close to the single generation case.

Beside being a very strong hypothesis about the UV completion of the SM, a drawback of the MFV hypothesis is that 
it does not allow us to define a clear power-counting in the SMEFT. This is because one of the breaking term, namely $y_t$, 
or better the 33 entries  of $Y_u Y_u^\dagger$ and $Y_u^\dagger Y_u$ in the basis~(\ref{eq:d-basis}), is large.
It is therefore not obvious why one should not consider more powers of $Y_u$ in the counting of independent 
operators, as for instance done in the non-linear realizations proposed in~\cite{Feldmann:2008ja,Kagan:2009bn}. 
However, it is only $y_t$ that is large, not the other entries of  $Y_u$. 
The insertion of an arbitrary powers of $y_t$ triggers the following breaking pattern
\be
U(3)_q \otimes  U(3)_u  \stackrel{y_t}{\longrightarrow}    U(2)_q \otimes  U(2)_u \otimes U(1)_{q^3_L +t_R}~.
\label{eq:U3toU2_yt}
\ee
A similar breaking to $U(2)$ subgroups occurs if we allow the third generation Yukawa couplings of down quarks 
and charged leptons to be large (a possibility that naturally occurs in models with an extended Higgs sector).
This observation, together with the more general argument that the third generation of fermions might play a special role in 
extensions of the SM,
 naturally brings us to consider a smaller symmetry group acting only on the light fermion families,
that is what we discuss  next.


\section{The $U(2)^5$ symmetry}
\label{sec:U2}

The $U(2)^5$ symmetry is the subgroup of $U(3)^5$ that, by construction, distinguish the first two generations of fermions 
from the third one~\cite{Barbieri:2011ci,Barbieri:2012uh,Blankenburg:2012nx}.
 It provides a ``natural'' explanation of why 
third-generation Yukawa couplings are large (being allowed by the symmetry) and, contrary to the MFV case, 
it allows us to build an EFT  where all the breaking terms are small, offering a more precise power counting 
for the operators. 

Given a fermion species $\psi_f$ ($f=\ell,q,e,u,d$),  the first two generations form a doublet of one 
of the $U(2)$ subgroups, whereas $\psi^{3}_f$ transform as a singlet. 
The five independent flavour doublets are denoted $L,Q,E,U,D$ and the flavour symmetry is decomposed as 
\begin{align}
	U(2)^5 = U(2)_L \otimes U(2)_Q \otimes U(2)_E \otimes U(2)_U \otimes U(2)_D~.
\end{align}
A set of symmetry breaking terms able to reproduce the observed SM Yukawa couplings,
which is minimal both in terms of the number of independent spurions, as well as in their size, 
is given by~\cite{Barbieri:2011ci}
\begin{eqnarray}
 &	V_\ell \sim\left(2,1,1,1,1\right)~,	\qquad   V_q \sim\left(1,2,1,1,1\right)~,  & \nonumber \\
 &  \Delta_e \sim\left(2,1,\bar{2},1,1\right)~, \qquad  \Delta_u \sim\left(1,2,1,\bar{2},1\right)~, \qquad  \Delta_d \sim\left(1,2,1,1,\bar{2}\right)~. &
\end{eqnarray}
By construction, $V_{q,\ell}$ are complex two-vectors and $\Delta_{e,u,d}$ are complex $2\times 2$~matrices. In terms of these spurions, we can express the Yukawa matrices as
\begin{align}
	Y_e = y_\tau\left(\begin{matrix}
		\Delta_e	 & x_\tau V_\ell \\
		0			 & 1
	\end{matrix}\right), && Y_u = y_t\left(\begin{matrix}
		\Delta_u	 & x_t V_q \\
		0			 & 1
	\end{matrix}\right), && Y_d = y_b\left(\begin{matrix}
		\Delta_d	 & x_b V_q \\
		0			 & 1
	\end{matrix}\right),
	\label{eq:YU2_5}
\end{align}
where $y_{\tau,t,b}$ and $x_{\tau,t,b}$ are free complex parameters expected to be of order~$\mathcal{O}(1)$. 
Alternative breaking terms, and the embedding of $U(2)^5$ in $U(3)^5$, are discussed in Section~\ref{sect:beyondU2}.

\paragraph{Explicit form for the spurions.} 
As already pointed out in the MFV case, the spurions can appear with arbitrary powers both in the renormalizable ($d=4$) part of the Lagrangian 
and in the dimension-six effective operators. In this case, we redefine the fields such that the kinetic terms are canonically normalised and the 
Yukawa couplings assume the form in Eq.~(\ref{eq:YU2_5}). This condition unambiguously normalises the $\Delta$ spurions,
but it leaves an $\mathcal{O}(1)$ freedom in the normalisation of the $V$ spurions (encoded by $x_{\tau,t,b}$).

Using the residual $U(2)^5$ invariance, we can transform the spurions to the following 
explicit form
\begin{equation}
 \Vql =    e^{i \bar \phi_{q(\ell)}} \begin{pmatrix}0 \\   \epsql  \end{pmatrix}~,  \quad  
\Delta_e = O_e^\intercal\, \begin{pmatrix} \delp_e  & 0 \\  0 & \del_e \end{pmatrix} ~, \quad
  \Delta_u= U_u^\dagger   \begin{pmatrix} \delp_u  & 0 \\  0 & \del_u \end{pmatrix}~,  \quad  
\Delta_d= U_d^\dagger  \begin{pmatrix} \delp_d  & 0 \\  0 & \del_d \end{pmatrix} ~,
\end{equation}
The flavour basis where the spurions assume this form is what we define as {\em interaction basis} for the fermion fields 
in the $U(2)^5$ setup. Here 
 $O$ and $U$ represent $2\times2$ orthogonal  and complex unitary matrices, respectively
\begin{align}\label{eq:Uq}
O_{e}=
\begin{pmatrix}
c_{e} & s_{e}  \\
-s_{e}   & c_{e}
\end{pmatrix}\,,
\qquad 
U_{q}=
\begin{pmatrix}
c_q & s_q\,e^{i\alpha_q}\\
-s_q\,e^{-i\alpha_q} & c_q
\end{pmatrix}\,,
\end{align}
with $s_i\equiv\sin\theta_i$ and $c_i\equiv\cos\theta_i$. The 
$\eps_i$ and   $\del^{(\prime)}_i$ are small positive real parameters 
controlling the overall size of the spurions. 
From the observed hierarchies of the Yukawa couplings, we deduce
\begin{equation}
1 \gg \eps_i   \gg  \del_i      \gg \delp_i     > 0
\end{equation}
or, more precisely,
\bea
&& \eps_i = \cO\left( {\rm Tr}(Y_u Y_u^\dagger) - \frac{ {\rm Tr}(Y_u Y_u^\dagger Y_d Y_d^\dagger)}{{\rm Tr}(Y_d Y_d^\dagger)} \right)^{1/2}
= \cO( y_t |V_{ts}| ) = \cO(10^{-1})~,  \\
&& 
\del_i  =  \cO\left( \frac{y_c}{y_t},   \frac{y_s}{y_b},   \frac{y_\mu}{y_\tau}  \right) = \cO(10^{-2})~,  \\ 
&& \delp_i  =  \cO\left( \frac{y_u}{y_t},   \frac{y_d}{y_b},   \frac{y_e}{y_\tau}  \right)   = \cO(10^{-3})~.
\label{eq:U2SpSize}
\eea
Starting from the interaction basis, the Yukawa couplings in~\eqref{eq:YU2_5} are diagonalized by unitary transformations
of the type $L_f^\dagger Y_f R_f  = {\rm diag}(Y_f)$, with $f=u,d,e$. The explicit 
form of these matrices is reported in Appendix~\ref{sect:appA}. While the  $\del^{(\prime)}_i$ are in one-to-one
correspondence with the light Yukawa eigenvalues, not all the other parameters appearing in the Yukawa and spurion decompositions
in Eqs.~(\ref{eq:YU2_5})--(\ref{eq:Uq}) can be put in correspondence with SM parameters (in particular with CKM elements). 
Contrary to the MFV case, 
in the $U(2)^5$ setup the structure of the spurions is not completely determined in terms of known parameters.
However, once we impose the hierarchy among the size of the spurions in Eq.~(\ref{eq:U2SpSize}), we effectively ``protect"
quark mixing as in the MFV case~\cite{Barbieri:2011ci}.

\subsection{Fermion bilinears}
We can now proceed classifying the number of independent operators appearing 
at $d=6$ in the SMEFT with a $U(2)^5$ flavour symmetry, minimally broken as discussed above.
Our final goal is to classify the operators up to 
$\mathcal{O}(V^3,\Delta^1 V^1)$, namely with up to three $V$ spurions (but no $\Delta$ terms), 
or with one $\Delta$ and at most one $V$. Given the size of the spurions in Eq.~(\ref{eq:U2SpSize}), 
this corresponds to neglecting terms which are at most of $\cO(10^{-4})$ according to our main hypotheses.

We start the analysis from the operators of classes 5, 6 and 7, which contains a fermion bilinear. 
To better illustrate how the hypothesis of a minimally broken $U(2)^5$ symmetry acts on the different 
 flavour structures, in the case of  left-handed and right-handed bilinears we analyse also the effect
of subleading breaking terms up to $\cO(\Delta^2 V^2)$.
More precisely, in the following we analyse how to span the flavour structure of the independent fermion bilinears 
in terms of the $U(2)^5$ breaking spurions. 


\paragraph{Left-handed  bilinears.}
As a representative example of left-handed fermion bilinears
we discuss in detail the  leptonic case (the translation to the quark case being trivial). For simplicity we omit $SU(2)_L$ and spinor indices, 
and often also flavour indices (except in expressions which would be ambiguous otherwise). The possible terms at different orders in the spurions 
for the case at hand is shown in Table~\ref{tab:LL}. The results can be summarised as follows in terms of the flavour tensor $\Lambda_{LL}$:
\be
\bar \ell_p  \Gamma \Lambda_{LL}^{pr} \ell_r~, \qquad 
\Lambda_{LL} =
	\left(\begin{matrix}
		a_1	& 0				& 0			\\
		0	& a_1 + c_1 \epsl^2		& \beta_1 \epsl	\\
		0	& \beta_1^\ast	\epsl & a_2		\\
	\end{matrix}\right) + \cO( \del_e^2)~.
	\label{eq:LambdaLL}
\ee
The explicit expression of $\Lambda_{LL}$ in Eq.~(\ref{eq:LambdaLL})
 corresponds to the expansion truncated at $\mathcal{O}(\Delta V)$ in the interaction basis. As can be seen,
at this order there is no mixing between the first generation and the others:\footnote{~Note that this statement holds only in the  interaction basis.}
$\Lambda_{LL}$, that in absence of any flavour symmetry is parameterised by 6 real and 3 imaginary coefficients
has only 4 real ($a_{1,2}$, $c_1$, Re$ \beta_1$) 
and 1 imaginary (Im$\beta_1$) coefficients.
A complete span of the whole $3\times3$ hermitian structure of  $\Lambda_{LL}$ 
occurs only with the inclusion of the terms up to $\cO(\Delta^2 V^2)$ shown in the lower part of Table~\ref{tab:LL}.

\renewcommand{\arraystretch}{1.4}
\begin{table}[t]
	\centering
	\begin{tabular}{l|l | l}
		Spurions 						& Operator & Explicit expression in flavour components \\ \hline
		$V^0$			& $a_1 \bar{L}L$  + $a_2 \bar{\ell}_3\ell_3$ & $a_1\left(\bar{\ell}_1\ell_1+\bar{\ell}_2\ell_2\right) +a_2 \left(\bar{\ell}_3\ell_3\right)$ \\
		$V^1$ 			& $\beta_1\bar{L}V_\ell \ell_3$~+~h.c. & $\beta_1 \epsl \left(\bar{\ell}_2 \ell_3\right)$~+~h.c. \\
		$V^2$ 			& $c_1 \bar{L}V_\ell V_\ell^\dagger L$ & $c_1  \epsl^2 \left(\bar{\ell}_2 \ell_2\right)$ \\
		$\Delta^1$,  	$\Delta^1 V^1$	  & -- &  -- \\ \hline
		$\Delta^2$ 		& $ h_1 \bar{L} \Delta_e \Delta_e^\dagger L $ & $ \approx h_1 \left[  \del_e^2  (\bar{\ell}_2 \ell_2)  -  s_e \del_e^2  (\bar{\ell}_1 \ell_2 +  \bar{\ell}_2 \ell_1) + (s_e^2 \del_e^2 + \delpsq_e )(\bar{\ell}_1 \ell_1)  \right]$ \\
		$\Delta^2 V^1$ 	& $\lambda_1 \bar{L}\Delta_e \Delta_e^\dagger V_\ell \ell_3$~+~h.c. & $\approx \lambda_1 \epsl \del_e^2  (  \bar{\ell}_2 \ell_3 
		-s_e \bar{\ell}_1 \ell_3 )$~+~h.c. \\
		$\Delta^2 V^2$	& $\mu_1 \bar{L}\Delta_e \Delta_e^\dagger V_\ell V_\ell^\dagger L$~+~h.c. & $\approx \mu_1  \epsl^2 \del_e^2 ( \bar{\ell}_2 \ell_2 - s_e 
		\bar{\ell}_1 \ell_2 )$~+~h.c.
	\end{tabular}
	\caption{Left-handed fermion bilinears allowed by different $U(2)$ breaking terms. The terms below the horizontal line are 
	subleading structures which are not considered in the general  analysis of independent terms.  
	The expressions in the third column are expanded 
	 in powers of $s_e$  up to first non-vanishing terms. 
	 \label{tab:LL} }
\end{table}

Here and in the following, when presenting explicit expressions, the phases of non-hermitian spurion combinations 
are reabsorbed into that of the corresponding complex coefficients. 
The criteria used to label the different terms are as follows: we denote with latin (greek) letters the real (complex) couplings appearing 
in hermitian (non-hermitian) structures. Terms with the same number of spurions are denoted with the same latin or greek letter
and different subscript.
Note that this notation focuses only on the flavour indices 
and not on the electroweak structure. A complete notation for the coupling of each operator 
can be chosen of the type $C_X ( F)$, where
$X$ denotes a specific electroweak structure, as in Table~\ref{tab:OperatorClasses}  ($X=H\ell, Hq,  \ldots$), 
and  $F=a_i, \beta_i, \ldots$ denotes the flavour structure.


\paragraph{Right-handed  bilinears.}
Proceeding in a similar manner, in Table~\ref{tab:RR} we report right-handed fermion bilinears  which are 
allowed by different spurion combinations. The leptonic bilinear $\bar e e$ is representative of any right-handed fermion bilinear with identical fields, 
while we treated separately the $\bar u d$ case which appears only for the operator $Q_{Hud}$. 
As far as identical fermions are concerned, we can express the result via the flavour tensor $\Lambda_{RR}$:
\be
\bar e_p  \Gamma \Lambda_{RR}^{pr} e_r~, \qquad 
\Lambda_{RR} =	
\left(\begin{matrix}
		a_1	& 0					&  \sigma_1^\ast \epsl  s_e \delp_e 	\\
		0	& a_1 		& \sigma_1^\ast \epsl  \del_e	\\
		\sigma_1 \epsl  s_e \delp_e	& \sigma_1 \epsl  \del_e	& a_2
	\end{matrix}\right)+ \cO(\del_e^2  )~.
\ee
Terminating the expansion up to $\cO(\Delta V)$, $\Lambda_{RR}$ contains 3 real and 1 imaginary coefficients.
At the same order, in the case of the (non-hermitian)  $\bar u d$ bilinear one finds 3 real and 3 imaginary coefficients
(see Table~\ref{tab:RR}).

Interestingly,  this structure  is quite ``robust'' with respect to higher-order corrections.
At  $\cO(\Delta^2)$ one generates a difference 
between the 11 and 22 entires of  $\Lambda_{RR}$, and only at $\cO(\Delta^2 V^2)$ non-vanishing 
12 and 21 entries, but this is not enough to span the entire  $3\times3$ hermitian structure:
this goal can be achieved only with inclusion of $\cO(\Delta^4 V^2)$ terms.
Most important, mixing terms involving first and/or second generations always require a suppression 
factor proportional to $\del_f$ and/or $\delp_f$. This is a feature related to 
our minimal choice of breaking terms.

\begin{table}[t]
	\centering
	\begin{tabular}{l|l | l}
		Spurions 						& Operator ($\bar e e$ type) & Explicit expression in flavour components \\ \hline
		$V^0$ 			& $a_1 \bar{E}E   +  a_2 \bar{e}_3 e_3 $  & $a_1\left(\bar{e}_1 e_1+\bar{e}_2 e_2\right)+a_2 \left(\bar{e}_3 e_3\right)$ \\
		$V^1,  V^2, 	\Delta^1$ 		         & -- & \\
		$\Delta^1 V^1$ 		& $ \sigma_1 \bar{e}_3 V_\ell^\dagger \Delta_e E$~+~h.c. & $ \approx  \sigma_1 \epsl   \left[ \del_e (\bar{e}_3 e_2)  + 
		s_e  \delp_e (\bar{e}_3 e_1) \right] $~+~h.c. \\ \hline
		$\Delta^2$ 		& $h_1 \bar{E} \Delta_e^\dagger \Delta_e E $ & 
								$  h_1  \left[  \del_e^2 ( \bar{e}_2 e_2)  + \delpsq_e( \bar{e}_1 e_1 ) \right]$ \\   
		$\Delta^2 V^1$ 	& -- &  \\
		$\Delta^2 V^2$	& $m_1 \bar{E}\Delta^\dagger_e  V_\ell V_\ell^\dagger \Delta_e  E$ & $\approx m_1  \epsl^2 \left[ 
		\del_e^2 ( \bar{e}_2 e_2) + s_e \delp_e \del_e ( \bar{e}_1 e_2+ \bar{e}_2 e_1)+s^2_e \delpsq_e ( \bar{e}_1 e_1 )  \right]$  \\
		\multicolumn{3}{c}{}  \\						
			Spurions 						& Operator ($\bar u d$ type) & Explicit expression in flavour components \\ \hline
		$V^0$ 		& $\alpha_1 \bar{u}_3 d_3$~+~h.c. & $ \alpha_1\left(\bar{u}_3 d_3\right)$~+~h.c. \\
				$V^1,  V^2, 	\Delta^1$ 		         & -- & \\
		$\Delta^1 V^1$  		& $\sigma_1 \bar{U} \Delta_u^\dagger V_q d_3$~+~h.c.
					& $ \approx  \sigma_1   \epsq  \left[   \del_u  \left(\bar{u}_2 d_3\right)  + s_u  e^{i \alpha_u} \delp_u   (\bar{u}_1 d_3)   \right]$~+~h.c. \\
		$\Delta^1 V^1$ 		& $\sigma_2 \bar{u}_3 V_q^\dagger \Delta_d D$~+~h.c.
					& $ \approx \sigma_2  \epsq  \left[  \del_d    \left(\bar{u}_3 d_2\right) + s_d    e^{-i \alpha_d} \delp_d  (\bar{u}_3 d_1)   \right]  $~+~h.c.
	\end{tabular}

	\caption{Right-handed fermion bilinears allowed by different $U(2)$ breaking terms. Notation as in Table~\ref{tab:LL}. }
	\label{tab:RR}
\end{table}

\paragraph{Left-right bilinears.}
The independent flavour structures of left-right fermion bilinear are listed in Table~\ref{tab:LR}, where we focus on the leptonic sector
as representative example.  Expressing the result via the flavour tensor $\Lambda_{LR}$ we find 
\be
\bar \ell_p  \Gamma \Lambda_{LR}^{pr} e_r~, \qquad 
\Lambda_{LR} =	
\left(\begin{matrix}
		 \rho_1  \delp_e 	& -  \rho_1 s_e \del_e 			& 0			\\
		 \rho_1  s_e \delp_e 	& \rho_1  \del_e 	& \beta_1 \epsilon_\ell	\\
		\sigma_1 \epsl  s_e \delp_e 	& \sigma_1 \epsl  \del_e 				& \alpha_1
	\end{matrix}\right)~+ \cO(\del_e \epsl^2  )~.
\ee
Terminating the expansion up to $\cO(\Delta V)$, we find 4 complex coefficients, to be compared with 
the potential 9 complex coefficients in absence of any flavour symmetry. For the same argument discussed in the case of 
the right-handed structures, in this case a span of the entire flavour space require terms with up to three powers of $\Delta$.

\begin{table}[t]
	\centering
	\begin{tabular}{l|l | l}
		Spurions 						& Operator & Explicit expression in flavour components \\ \hline
		$V^0$ 			& $\alpha_1 \bar{\ell}_3 e_3  $ 		& $\alpha_1 \left(\bar{\ell}_3 e_3\right)  $ \\
		$V^1$ 			& $\beta_1 \bar{L}V_\ell e_3  $ & $\beta_1 \epsl \left(\bar{\ell}_2 e_3\right)  $ \\
		$V^2$ 			& -- & \\
		$\Delta^1$ 		& $\rho_1 \bar{L} \Delta_e E  $ & 
		$ \approx  \rho_1   \left[  \delta_e \left(\bar{\ell}_2 e_2\right) -  s_e \delta_e \left(\bar{\ell}_1 e_2\right) +  s_e  \delp_e \left(\bar{\ell}_2 e_1\right) + 		
		\delp_e \left(\bar{\ell}_1 e_1\right) \right]$ \\
		$\Delta^1 V^1$ 		& $\sigma_1 \bar{\ell}_3 V_\ell^\dagger \Delta_e E $ & 
		$\approx  \sigma_1  \epsl   \left[  \delta_e \left(\bar{\ell}_3 e_2\right)  + s_e \delp_e \left(\bar{\ell}_3 e_1\right)   \right] $
	\end{tabular}
	\caption{Left-right fermion bilinears allowed by different $U(2)$ breaking terms (the sum over hermitian conjugates is understood for all structures).
	Notation as in Table~\ref{tab:LL}. 
}
	\label{tab:LR}
\end{table}


\paragraph{Summary.}
The total number of CP-even and CP-odd coefficients for all the operators with fermion bilinears constructed with spurions up to $\mathcal{O}(\Delta^1 V^1)$ 
are reported in Table~\ref{tab:2FermionResult}.
\begin{table}[t]
	\centering
	\renewcommand{\arraystretch}{1.2} 
	\begin{tabular}{l | c|| l l|l l|l l|l l|l l}
		 &  N.~indep. & \multicolumn{10}{c}{ $U(2)^5$ breaking terms} \\ 
		 			Class & structures & \multicolumn{2}{c|}{$V^0$} & \multicolumn{2}{c|}{$V^1$} & \multicolumn{2}{c|}{$V^2$} & \multicolumn{2}{c|}{$\Delta^1$} & \multicolumn{2}{c}{$\Delta^1V^1$} \\ \hline
		5 \& 6: $\left( \bar{L}R \right)$ 	& 11 			& 11 		& 11 	& 11 	& 11 	& -- 	& -- 	& 11 	& 11 	& 11 	& 11 \\
		7: $\left( \bar{L}L \right)$			& 4 			& 8 		& -- 	& 4 	& 4 	& 4 	& -- 	& -- 	& -- 	& -- 	& -- \\
		7: $\left( \bar{R}R \right)$			& 3 			& 6 		& -- 	& -- 	& -- 	& -- 	& -- 	& -- 	& -- 	& 3 	& 3 \\
		7: $Q_{Hud}$					& 1 			& 1 		& 1 	& -- 	& -- 	& -- 	& -- 	& -- 	& -- 	& 2 	& 2 \\ \hline
		total:								& 19			& 26 	& 12 	& 15	& 15	& 4 	& -- 	& 11	& 11	& 16 	& 16
	\end{tabular}
	\caption{Number of independent operators with fermion bilinears in $U(2)^5$. Notation as in Table~\ref{tab:U3new};
	 however, here each column denotes the operators with a precise power of spurions, as indicated in the first row. 
	 	\label{tab:2FermionResult}}
\end{table}

\subsection{Four fermion operators.}
In this section we proceed analysing the operators in class~8 which contain four fermion fields. 
In analogy to the 2-index tensors $\Lambda$ introduced to describe the fermion bilinears,  the
flavour structure of theses operators is  described by 4-index tensors $\Sigma$. As an illustration, 
and also in view of the phenomenological application in Sect.~\ref{sect:pheno},
in the case of $(\bar{L}L)(\bar{L}L)$ operators we present the explicit component 
structure of these tensor. For the other operators we simply list the allowed structures up to 
$\cO(V^3,\Delta^1 V^1)$.

\paragraph{$(\bar{L}L)(\bar{L}L)$ structures.}
In this category of  operators we can distinguish two different subclasses as far as flavour structure and spurion analysis are concerned. The first one
contains operators where both bilinears are of the same form, namely $Q_{\ell\ell}$, $Q_{qq}^{(1)}$ and $Q_{qq}^{(3)}$. Considering $Q_{\ell\ell}$ as representative example
of this class of operators, the terms generated up to $\cO(V^3)$ are
\be
\begin{array}{ll}
V^0 :		& \big[ a_1 (\bar{L}^p L^p )(\bar{L}^r L^r)
		+ a_2 (\bar{L}^p L^r)(\bar{L}^r L^p)
		+ a_3 (\bar{L}L)(\bar{\ell}_3 \ell_3)  \\
		& + a_4 (\bar{L}\ell_3)(\bar{\ell}_3 L)
		+ a_5 (\bar{\ell}_3\ell_3)(\bar{\ell}_3 \ell_3) \big]~,  \\
V^1:		&  \big[ \beta_1 (\bar{L}^p V_\ell^p \ell_3)(\bar{L}^r L^r) 
		+ \beta_2 (\bar{L}V_\ell \ell_3)(\bar{\ell}_3 \ell_3) 
		+ \beta_3 (\bar{L}^p V_\ell^p L^r)(\bar{L}^r \ell_3)~+~{\rm h.c.} \big]~, \\ 
V^2:		& \big[ c_1 (\bar{L}^p V_\ell^p V_\ell^{\dagger\,r} L^r)(\bar{L}^s L^s) 
		+ c_2 (\bar{L}^p V_\ell^p V_\ell^{\dagger\,r} L^r)(\bar{\ell}_3 \ell_3)	
		+ c_3 (\bar{L}^p V_\ell^p \ell_3)(\bar{\ell}_3 V_\ell^{\dagger\,r} L^r) \\ &
		+ c_4 (\bar{L}^p V_\ell^p L^r)(\bar{L}^r V_\ell^{\dagger\,s} L^s) 
		+ (\gamma_1 (\bar{L}^p V_\ell^p \ell_3)(\bar{L}^r V_\ell^r \ell_3)~+~{\rm h.c.}) \big]~,	 \\
V^3: 		& \big[ \xi_1 (\bar{L}^p V_\ell^p V_\ell^{\dagger\,r} L^r)(\bar{L}^s V_\ell^s \ell_3)~+~{\rm h.c.}\big]~.
\end{array}
\label{eq:4La}
\ee
For the remaining two operators, $Q_{\ell q}^{(1)}$ and $Q_{\ell q}^{(3)}$, we get the following terms:
\be
\begin{array}{ll} 
V^0 :		& \big[a_1 (\bar{L}L)(\bar{Q}Q) 	
		+ a_2 (\bar{L}L)(\bar{q}_3 q_3)	
		+ a_3 (\bar{\ell}_3 \ell_3)(\bar{Q} Q)  
		+ a_4 (\bar{\ell}_3\ell_3)(\bar{q}_3 q_3)\big]~,  \\
V^1:		&\big[ \beta_1 (\bar{L}V_\ell \ell_3)(\bar{Q}Q) 
		+ \beta_2 (\bar{L}V_\ell \ell_3)(\bar{q}_3 q_3)
		+ \beta_3 (\bar{L}L)(\bar{Q}V_q q_3) 
 		+ \beta_4 (\bar{\ell}_3 \ell_3)(\bar{Q}V_q q_3)~+~\text{h.c.}\big]~,	 \\
V^2:		& \big[c_1 (\bar{L}^p V_\ell^p V_\ell^{\dagger\,r} L^r)(\bar{Q}Q) 
		+ c_2 (\bar{L}^p V_\ell^p V_\ell^{\dagger\,r} L^r)(\bar{q}_3 q_3)
		+ c_3 (\bar{L} L)(\bar{Q}^p V_q^p V_q^{\dagger\,r} Q^r) \\ &
		+ c_4 (\bar{\ell}_3 \ell_3)(\bar{Q}^p V_q^p V_q^{\dagger\,r} Q^r) 
		+ (\gamma_1 (\bar{L} V_\ell \ell_3)(\bar{Q} V_q q_3)
		+ \gamma_2 (\bar{L}V_\ell \ell_3)(\bar{q}_3 V_q^\dagger Q)~+~\text{h.c.})\big]~, \\
V^3:		&\big[ \xi_1 (\bar{L}^p V_\ell^p V_\ell^{\dagger\,r} L^r)(\bar{Q} V_q q_3)
		+ \xi_2 (\bar{L}V_\ell \ell_3)(\bar{Q}^p V_q^p V_q^{\dagger\,r} Q^r)~+~\text{h.c.}\big]~.
\end{array}
\label{eq:4Lb}
\ee
No additional terms arise with the insertion of one power of $\Delta$. On the other hand, it is worth stressing that 
the $(\bar{L}L)(\bar{L}L)$ operators are the only ones where terms with 3 powers of the $V$ spurions are relevant
(more details about the number of independent fermion contractions for four-fermion operators are given in Appendix~\ref{app:contractions}).

For each electroweak structure of  $(\bar{L}L)(\bar{L}L)$ operators  we therefore find the following number of real and imaginary coefficients  
at a given order in the spurion expansion:
\be
	\begin{array}{l  l l|l l|l l|l l|l l|l l}
	&	\multicolumn{2}{c|}{V^0} & \multicolumn{2}{c|}{V^1} & \multicolumn{2}{c|}{V^2} & \multicolumn{2}{c|}{\Delta^1} & \multicolumn{2}{c|}{\Delta^1 V^1} & \multicolumn{2}{c}{V^3} \\ \cline{2-13}
{\rm Type~``a"}~[Q_{\ell \ell}, Q_{qq}^{(1,3)}]:	 &	5 & - & 3 & 3 & 5 & 1 & - & - & - & - & 1 & 1  \\   
{\rm Type~``b"}~[Q_{\ell q}^{(1)}, Q_{\ell q}^{(3)}]:	 &	4 & - & 4 & 4 & 6 & 2 & - & - & - & - & 2 & 2
	\end{array}
\ee
As anticipated,  the flavour structure of the four-fermion operators is described by the 4-index tensors $\Sigma$. In the specific case 
of the structures in Eqs.~(\ref{eq:4La})--(\ref{eq:4Lb}) we defined them as
\be
  \Sigma_{\ell \ell}^{ij,nm}  \left(\bar{\ell}_i  \Gamma \ell_j\right) \left(\bar{\ell}_n \Gamma \ell_m\right)  
  \qquad {\rm and} \qquad 
  \Sigma_{ \ell q}^{ij,nm}  \left(\bar{\ell}_i  \Gamma \ell_j\right) \left(\bar{q}_n \Gamma q_m\right)~.
  \label{eq:SigmaTens}
\ee
The corresponding explicit expressions are reported in Table~\ref{tab:SigmaLL} and \ref{tab:SigmaQL} 
in Appendix~\ref{app:tables}, respectively.

\paragraph{$(\bar{R}R)(\bar{R}R)$ structures.}
In this case we can distinguish three different subclasses of operators.
The first one includes operators with identical right-handed quark fields, namely $Q_{uu}$ and $Q_{dd}$. 
Considering $Q_{uu}$  as representative example of this subclass of operators, the terms generated up to $\cO(V^3, \Delta^1V^1)$ are
\be
\begin{array}{ll}
V^0 :		& \big[ a_1 (\bar{U}^p U^p)(\bar{U}^r U^r) 	
		+ a_2 (\bar{U}^p U^r)(\bar{U}^r U^p) 	
		+ a_3 (\bar{U}U)(\bar{u}_3 u_3) \\
		& + a_4 (\bar{U}u_3)(\bar{u}_3 U)	
		+ a_5 (\bar{u}_3 u_3)(\bar{u}_3 u_3)\big]~, \\
\Delta^1 V^1:	& \big[ \sigma_1(\bar{u}_3 V_q^{\dagger\,s} \Delta_u^{s,r} U^r)(\bar{U}^p U^p)
		+ \sigma_2 (\bar{u}_3 V_q^\dagger \Delta_u U)(\bar{u}_3 u_3)  \\
		&  + \sigma_3 (\bar{U}^p V_q^{\dagger\,s} \Delta_u^{sr} U^r)(\bar{u}_3 U^p)~+~{\rm h.c.}  	 \big]~.
\end{array}
\label{eq:4Ra}
\ee
The second type is the operator $Q_{ee}$, which also involves identical right-handed fields. The decomposition proceeds as for $Q_{uu}$; 
however,  due to the Fierz identity in Eq.~(\ref{eq:Fierz}), 
we should not consider as independent terms of the type $(\bar{E}e_3)(\bar{e}_3 E)$ and $(\bar{E}^p E^r)(\bar{E}^r E^p)$, which reduce to $(\bar{E}E)(\bar{e}_3 e_3)$ and $(\bar{E}^p E^p)(\bar{E}^r E^r)$ respectively.  
Similarly, at higher order in the spurion expansion, we can relate the operator 
$( V_q^{\dagger\,s} \Delta_e^{s,r} ) (\bar{E}^p E^r)(\bar{e}_3 E^p)$ to $( V_q^{\dagger\,s} \Delta_e^{s,r} ) (\bar{E}^p E^p)(\bar{e}_3 E^r)$.

For the remaining four operators $Q_{eu}$, $Q_{ed}$, $Q_{ud}^{(1)}$ and $Q_{ud}^{(8)}$ the counting is the same\footnote{~This statement holds because 
we truncate the spurion expansion up to $\cO(V^3, \Delta^1V^1)$: at higher orders the counting for $Q_{eu}$ and $Q_{ed}$ would start to differ from the counting for $Q_{ud}^{(1)}$ and $Q_{ud}^{(8)}$.}. Considering $Q_{eu}$ as representative example of this subclass we find
\be
\begin{array}{ll}
V^0:		& \big[ a_1 (\bar{E}E)(\bar{U}U) 		
		+ a_2 (\bar{E}E)(\bar{u}_3 u_3)	
		+ a_3 (\bar{e}_3 e_3)(\bar{U}U)	
		+ a_4 (\bar{e}_3 e_3)(\bar{u}_3 u_3)\big] ~, \\ 
\Delta^1 V^1:	& \big[ \sigma_1(\bar{e}_3 V_\ell^\dagger \Delta_e E)(\bar{U}U) 	
		+ \sigma_2 (\bar{e}_3 V_\ell^\dagger \Delta_e E)(\bar{u}_3 u_3)	 \\	
		&+  \sigma_3 (\bar{E}E)(\bar{u}_3 V_q^\dagger \Delta_u U) 	
		+ \sigma_4 (\bar{e}_3 e_3)(\bar{u}_3 V_q^\dagger \Delta_u U) ~+~{\rm h.c.}  \big]~.
\end{array}
\label{eq:4Rb}
\ee		
For each electroweak structure of  $(\bar{R}R)(\bar{R}R)$ operators  we therefore find the following number of real and imaginary coefficients  
at a given order in the spurion expansion:
\be
	\begin{array}{l l l|l l|l l|l l|l l|l l}
		& \multicolumn{2}{c|}{V^0} & \multicolumn{2}{c|}{V^1} & \multicolumn{2}{c|}{V^2} & \multicolumn{2}{c|}{\Delta^1} & 
		\multicolumn{2}{c|}{\Delta^1 V^1} & \multicolumn{2}{c}{V^3} \\ \cline{2-13}
		{\rm Type~``a_1"}~[Q_{uu(dd)}]:	& 5 & - & - & - & - & - & - & - & 3 & 3 & - & -  \\
		{\rm Type~``a_2"}~[Q_{ee}]: 			& 3 & - & - & - & - & - & - & - & 2 & 2 & - & - \\
	{\rm Type~``b"}~[Q_{eu}, Q_{ed}, Q_{ud}^{(1,8)}]:  &	4 & - & - & - & - & - & - & - & 4 & 4 & - & -\\
	\end{array}
\ee


\paragraph{$(\bar{L}L)(\bar{R}R)$ structures.}
In this case we distinguish again two sub-classes. The first one includes $Q_{le}$, $Q_{qu}^{(1,8)}$,  and $Q_{qd}^{(1,8)}$.
Considering $Q_{le}$ as representative example, the terms generated up to $\cO(V^3, \Delta^1V^1)$ are
\be
\begin{array}{ll}
V^0:		& \big[ a_1 (\bar{L}L)(\bar{E}E)	
		+  a_2 (\bar{L}L)(\bar{e}_3 e_3)	
		+ a_3 (\bar{\ell}_3 \ell_3)(\bar{E}E)	
		+ a_4 (\bar{\ell}_3 \ell_3)(\bar{e}_3 e_3)\big]~,   \\
V^1:		& \big[ \beta_1 (\bar{L} V_\ell \ell_3)(\bar{E}E)	
		+ \beta_2 (\bar{L} V_\ell \ell_3)(\bar{e}_3 e_3)	~+~{\rm h.c.}  \big]~,\\
V^2:		& \big[ c_1 (\bar{L}^p V_\ell^p V_\ell^{\dagger\,r} L^r)(\bar{E}E)
		+ c_2 (\bar{L}^p V_\ell^p V_\ell^{\dagger\,r} L^r)(\bar{e}_3 e_3)\big]~, \\
\Delta^1 V^0:	& \big[ \rho_1(\bar{L}\ell_3)\Delta_e(\bar{e}_3 E) ~+~{\rm h.c.}  \big]~,\\
\Delta^1 V^1:	&\big[ \sigma_1(\bar{L}^p V_\ell^{\dagger \, r} L^r)\Delta_e^{pt}(\bar{e}_3 E^t)
		+ \sigma_2 (\bar{L}^p L^p)V_\ell^{\dagger\,r} \Delta_e^{rt}(\bar{e}_3 E^t)  \\
		& +   \sigma_3 (\bar{\ell}_3 \ell_3)V_\ell^\dagger \Delta_e(\bar{e}_3 E) ~+~{\rm h.c.}  \big]~.
\end{array}
\label{eq:4LRa}
\ee	
For the remaining operators $Q_{lu}$, $Q_{ld}$, and $Q_{qe}$, considering $Q_{lu}$ as representative, we get 
\be
\begin{array}{ll}
V^0:		& \big[ a_1  (\bar{L}L)(\bar{U}U)			
 		+ a_2 (\bar{L}L)(\bar{u}_3 u_3)	
  		+ a_3 (\bar{\ell}_3 \ell_3)(\bar{U}U)	
  	 	+ a_4 (\bar{\ell}_3 \ell_3)(\bar{u}_3 u_3)\big]~, \\
V^1:	 	& \big[  \beta_1 (\bar{L} V_\ell \ell_3)(\bar{U}U)
	 	+ \beta_2 (\bar{L} V_\ell \ell_3)(\bar{u}_3 u_3)~+~{\rm h.c.}  \big]~,\\
V^2:		& \big[ c_1 (\bar{L}^p V_\ell^p V_\ell^{\dagger\,r} L^r)(\bar{U}U) 	
 		+ c_2 (\bar{L}^p V_\ell^p V_\ell^{\dagger\,r} L^r)(\bar{u}_3 u_3)\big]~, \\
\Delta^1 V^1:	& \big[ \sigma_1(\bar{L} L)V_q^\dagger \Delta_u(\bar{u}_3 U)	
		+ \sigma_2 (\bar{\ell}_3 \ell_3)V_q^\dagger \Delta_u(\bar{u}_3 U)~+~{\rm h.c.}  \big]~. 
\end{array}
\label{eq:4LRb}
\ee
For each electroweak structure of  $(\bar{L}L)(\bar{R}R)$ operators  we therefore find the following number of real and imaginary coefficients  
at a given order in the spurion expansion:
\be
	\begin{array}{l l l|l l|l l|l l|l l|l l}
		& \multicolumn{2}{c|}{V^0} & \multicolumn{2}{c|}{V^1} & \multicolumn{2}{c|}{V^2} & \multicolumn{2}{c|}{\Delta^1} & 
		\multicolumn{2}{c|}{\Delta^1 V^1} & \multicolumn{2}{c}{V^3} \\ \cline{2-13}
	{\rm Type~``a"}~[Q_{le}, Q_{qu}^{(1,8)},  Q_{qd}^{(1,8)}]: &	4 & - & 2 & 2 & 2 & - & 1 & 1 & 3 & 3 & - & - \\
	{\rm Type~``b"}~[Q_{lu}, Q_{ld}, Q_{qe}]:  	& 				4 & - & 2 & 2 & 2 & - & - & - & 2 & 2 & - & -
	\end{array}
\ee

\paragraph{$(\bar{L}R)(\bar{R}L)$ and $(\bar{L}R)(\bar{L}R)$ structures.}
There is a single $(\bar{L}R)(\bar{R}L)$ operator, $Q_{ledq}$, for which the spurion decomposition up to $\cO(V^3, \Delta^1V^1)$ 
yields:
\be
\begin{array}{ll}
V^0:		& \big[ \alpha_1 (\bar{\ell}_3 e_3)(\bar{d}_3 q_3) ~+~{\rm h.c.}  \big]~,		\\
V^1:	 	& \big[ \beta_1 (\bar{L} V_\ell e_3)(\bar{d}_3 q_3)	
		+ \beta_2 (\bar{\ell}_3 e_3)(\bar{d}_3 V_q^\dagger Q) ~+~{\rm h.c.}  \big]~, \\
V^2:		& \big[ \gamma_1 (\bar{L} V_\ell e_3)(\bar{d}_3 V_q^\dagger Q)  ~+~{\rm h.c.}  \big]~, \\
\Delta^1 V^0:	& \big[ \rho_1(\bar{L} \Delta_e E)(\bar{d}_3 q_3)
 		+ \rho_2 (\bar{\ell}_3 e_3)(\bar{D}\Delta_d^\dagger Q) ~+~{\rm h.c.}  \big]~, \\
\Delta^1 V^1: 	& \big[  \sigma_1(\bar{L} \Delta_e E)(\bar{d}_3 V_q^\dagger Q)
 		+ \sigma_2 (\bar{\ell}_3 V_\ell^\dagger \Delta_e E)(\bar{d}_3 q_3) 	\\
		& + \sigma_3 (\bar{L} V_\ell e_3)(\bar{D}\Delta_d^\dagger Q)		
		+ \sigma_4 (\bar{\ell}_3 e_3)(\bar{D}\Delta_d^\dagger V_q q_3) ~+~{\rm h.c.}  \big]~,
\end{array}
\label{eq:4LRc}
\ee
As far as $(\bar{L}R)(\bar{L}R)$ structures are concerned, we need to distinguish between $Q_{lequ}^{(1,3)}$ 
and $Q_{quqd}^{(1,3)}$. In the first case we have the same decomposition as for $Q_{ledq}$, while in the second case we get
\be
\begin{array}{ll}
V^0:		& \big[  \alpha_1 (\bar{q}_3 u_3)(\bar{q}_3 d_3) ~+~{\rm h.c.}  \big]~,  \\
 V^1:	 	& \big[ \beta_1 (\bar{Q} V_q u_3)(\bar{q}_3 d_3)	
  		+ \beta_2 (\bar{q}_3 u_3)(\bar{Q} V_q d)	~+~{\rm h.c.}  \big]~, \\
 V^2:	 	& \big[  \gamma_1 (\bar{Q} V_q u_3)(\bar{Q} V_q d_3)  ~+~{\rm h.c.}  \big]~,\\
 \Delta^1 V^0:	& \big[   \rho_1(\bar{Q} \Delta_u U)(\bar{q}_3 d_3)	+ \rho_2 (\bar{q}_3 u_3)(\bar{Q}\Delta_d D) \\
      		&+ \rho_3 (\bar{q}_3 \Delta_u U)(\bar{Q} d_3)+ \rho_4 (\bar{Q}u_3)(\bar{q}_3 \Delta_d D)	 ~+~{\rm h.c.}  \big]~,\\
 \Delta^1 V^1: 	& \big[   \sigma_1(\bar{q}_3 V_q^\dagger \Delta_u U)(\bar{q}_3 d_3) 	
         	+ \sigma_2 (\bar{Q}^p \Delta_u^{pr} U^r)(\bar{Q}^s V_q^s d_3) 	
          	+ \sigma_3 (\bar{q}_3 u_3)(\bar{q}_3 V_q^\dagger \Delta_d D) 	\\
         	& + \sigma_4 (\bar{Q}^p V_q^p u_3)(\bar{Q}^r \Delta_d^{rs} D^s)	
           	+ \sigma_5 (\bar{Q}^p V_q^p U^r)\Delta_u^{sr}(\bar{Q}^s d_3)
            	+ \sigma_6 (\bar{Q}^p u_3)\Delta_d^{pr}(\bar{Q}^s V_q^s D^r) ~+~{\rm h.c.}  \big]~.
\end{array}
\label{eq:4LRd}
\ee
For each electroweak structure of  $(\bar{L}R)(\bar{R}L)$ and $(\bar{L}R)(\bar{L}R)$ operators  we therefore find the following number of real and imaginary coefficients  at a given order in the spurion expansion:
\be
	\begin{array}{l l l|l l|l l|l l|l l|l l}
		& \multicolumn{2}{c|}{V^0} & \multicolumn{2}{c|}{V^1} & \multicolumn{2}{c|}{V^2} & \multicolumn{2}{c|}{\Delta^1} & 
		\multicolumn{2}{c|}{\Delta^1 V^1} & \multicolumn{2}{c}{V^3} \\ \cline{2-13}
	{\rm Typa~``a"}~[Q_{ledq}, Q_{lequ}^{(1,3)}]:  	&  1 & 1 & 2 & 2 & 1 & 1 & 2 & 2 & 4 & 4 & - & -  \\
	{\rm Type~``b"}~[Q_{quqd}^{(1,8)}]:  		 	& 1 & 1 & 2 & 2 & 1 & 1 & 4 & 4 & 6 & 6 & - & -
\end{array}
\label{eq:4LRd2}
\ee

\begin{table}[t]
\centering
	\renewcommand{\arraystretch}{1.2} 
	\begin{tabular}{ l||l l|l l|l l|l l|l l|l l|l l}
	&   \multicolumn{14}{c}{ $U(2)^5 \quad$   [terms summed up to different orders]}   \\
 Operators & \multicolumn{2}{c|}{Exact} & \multicolumn{2}{c|}{$\mathcal{O}(V^1)$} & \multicolumn{2}{c|}{$\mathcal{O}(V^2)$} & \multicolumn{2}{c|}{$\mathcal{O}(V^1,\Delta^1)$} & \multicolumn{2}{c|}{$\mathcal{O}(V^2,\Delta^1)$} & \multicolumn{2}{c|}{$\mathcal{O}(V^2,\Delta^1 V^1)$} & \multicolumn{2}{c}{$\mathcal{O}(V^3,\Delta^1 V^1)\!\!\!\! $} \\ \hline
Class 1--4						& 9 	& 6 	& 9 	& 6 	& 9 	& 6 	& 9 	& 6 	& 9 	& 6 	& 9 	& 6 	 & 9 	& 6 			\\ \hline
$\psi^2 H^3$ 					& 3 	& 3		& 6		& 6		& 6 	& 6  	& 9 	& 9		& 9		& 9		& 12	& 12	& 12	& 12 	\\
$\psi^2 X H$ 	 				& 8		& 8		& 16	& 16	& 16	& 16	& 24	& 24	& 24	& 24	& 32	& 32	& 32	& 32 	\\
$\psi^2 H^2 D$					& 15	& 1		& 19	& 5		& 23	& 5 	& 19	& 5		& 23	& 5		& 28	& 10	& 28	& 10 	\\ \hline
$(\bar{L}L)(\bar{L}L)$			& 23	& --		& 40	& 17	& 67	& 24	& 40	& 17	& 67	& 24	& 67	& 24	& 74	& 31 	\\
$(\bar{R}R)(\bar{R}R)$			& 29	& --		& 29	& --		& 29	& -- 	& 29	& -- 	& 29	& -- 	& 53	& 24	& 53	& 24 	\\
$(\bar{L}L)(\bar{R}R)$			& 32	& --		& 48	& 16	& 64	& 16	& 53	& 21	& 69	& 21	& 90	& 42	& 90	& 42 	\\
$(\bar{L}R)(\bar{R}L)$  			& 1		& 1		& 3		& 3		& 4		& 4 	& 5		& 5		& 6		& 6		& 10	& 10	& 10	& 10 	\\
$(\bar{L}R)(\bar{L}R)$ 			& 4		& 4		& 12	& 12	& 16	& 16	& 24	& 24	& 28	& 28	& 48	& 48	& 48	& 48 	\\  \hline
	\multicolumn{1}{c||}{\bf total:}					 	& 124	& 23	& 182	& 81	& 234	& 93	& 212	& 111	& 264	& 123	
	& 349	& 208	& 356	& 215	\\
	\end{tabular}
	\caption{Number of independent operators in the SMEFT assuming a minimally broken $U(2)^5$ symmetry, including 
	breaking terms up to $\cO(V^3, \Delta^1V^1)$. Notations as in Table~\ref{tab:U3new}.
        \label{tab: final results} }
\end{table}

\subsection{Summary and discussion}
The results for all SMEFT operators are summarized in Table~\ref{tab: final results}, while the detailed counting order by order,
organised according to the different sub-categories of operators is presented in  Table~\ref{tab:U2detailed}
in Appendix~\ref{app:tables}. As expected, the smaller symmetry group leads to a significantly larger number of terms compared
to the MFV case in Table~\ref{tab:U3new}. However, we emphasise that the number of independent terms is  still 
rather small compared to the case of no symmetry, even when considering high powers of the spurions. 
 It is also worth stressing that 
the smallness (and the nature) of the $U(2)^5$ breaking terms allows us to consider only limited subsets of the terms reported in 
Table~\ref{tab: final results} depending on the observables, and the level of precision, we are interested in. For instance, in the limit where
we neglect the masses of the first two generations (which is often an excellent approximation) we can stop at the third column.

\section{Beyond $U(3)^5$ and $U(2)^5$ with minimal breaking.}
\label{sect:beyondU2}

The main virtue of the MFV hypothesis is to normalise the magnitude of flavour-violating processes in the quark sector 
to their corresponding size in the SM (controlled by the CKM matrix), 
both for tree-level amplitudes and for the leading loop-induced ones. This is achieved by linking every possible source of flavour non-universality 
to the Yukawa couplings. Beside this rather strong assumption from the model-building point of view, the MFV hypothesis 
has two main drawbacks in the EFT implementation: i)~no clear power-counting 
due to the large value of $y_t$ (hence no special role for the third generation which, on the other hand, plays a key role in the Higgs 
hierarchy problem);~ii) no flavour mixing in the lepton sector.
The advantage of the $U(2)^5$ setup, with the minimal breaking discussed in Section~\ref{sec:U2}, 
is that it addresses these two drawbacks preserving the main virtue of the MFV setup in flavour-violating processes.
Moreover, this goal is achieved without imposing a  specific alignment among the spurions, but only a well-defined 
hierarchical structure.  The price to pay is a significant enlargement in the number of free parameters due to the smaller 
symmetry group. In this section we briefly discuss if it is worth to consider alternative options, 
either as far as flavour symmetries or as far as symmetry-breaking terms are concerned. 

\subsection{Unbroken $U(3)_{d,e}$ groups.} 
A first natural question to ask is if we can consider an intermediate case, combining $U(2)$ groups in the up-quark sector and $U(3)$ 
subgroups in the right-handed down-quark and/or charged-lepton sector.
To clarify this point, let's consider the breaking chain in the quark sector due to the third-generation Yukawa couplings. Similarly to 
Eq.~(\ref{eq:U3toU2_yt}), if we consider a two-step breaking due to $y_t$ and, later on by $y_b$ (that eventually we are interested to treat as a spurion), 
the impact on the whole quark-flavour symmetry group is~\cite{Feldmann:2008ja,Kagan:2009bn}\footnote{~Here 
$U(1)^{\prime}_{3}=U(1)_{q^3_L + t_R} $ and $U(1)^{\prime\prime}_{3}=U(1)_{q^3_L +b_R+ t_R}$.}
\be
U(3)_q   \otimes  U(3)_u  \otimes  U(3)_d  \stackrel{y_t}{\longrightarrow}   U(2)_q \otimes  U(2)_u \otimes   U(3)_d  \otimes   U(1)_3^\prime  
 \stackrel{ y_{t,b}}{\longrightarrow}   U(2)_q  \otimes  U(2)_u \otimes     U(2)_d  \otimes   U(1)^{\prime\prime}_{3}~.
\ee
As can be seen, in the first step we are left with an unbroken $U(3)$ subgroup.  However,
this case is less appealing than the $U(2)^5$ minimal setup, since we would need to impose a spefic 
alignment among the spurions to recover a MFV-like suppression of flavour-changing amplitudes.
This  follows from the decompositon of the SM Yukawa couplings
in terms of spurions of the unbroken subgroups in the two cases:
\bea
  Y_{u,d}  & \stackrel{y_t}{\longrightarrow}   &  V_q   \oplus \Delta_u  \oplus \left[   \Lambda_d \sim (1,1,\bar 3)  
   \oplus \Sigma_d \sim (2,1,\bar 3)  \oplus V^{u\phantom{d}}_R  \sim (1,\bar 2,1)  
   \right]_{U(2)_q  \otimes  U(2)_u \otimes     U(3)_d}~,
    \label{eq:U2U3} \\ 
   Y_{u,d}  & \stackrel{y_{t,b}}{\longrightarrow}   &   V_q  \oplus \Delta_u  \oplus \Delta_d  \oplus \left[  V^u_R  \sim (1,\bar 2,1)  \oplus V^d_R  \sim (1,1, \bar 2) \right]_{
   U(2)^3 }~.
   \label{eq:U2direct}
\eea
Here $ V_q$ and $\Delta_{u,d}$ denote the spurions that we have in the $U(2)^5$ minimal setup, whereas between brackets we show the 
additional terms arising in the decomposition. 
\begin{itemize}
\item[I]{\em No unbroken $U(3)$ subgroups,}~Eq.~(\ref{eq:U2direct}). Here 
 we simply need to assume the hierarchy  $ |V_q|  \gg  |\Delta_{u,d}| \gg |V^{u,d}_R| $ to recover a MFV-like structure
 and, at the same time, have all terms necessary to reconstruct completely the SM Yukawa couplings. 
\item[II]{\em Unbroken $U(3)_d$ group,}~Eq.~(\ref{eq:U2U3}). Here we cannot neglect the terms between brackets
if we wish to describe down-quark masses. Allowing only $\Lambda_d \not=0$ we can describe $y_b\not=0$, but 
strange and down quark remain massless. A description of all the masses require both $\Lambda_d \not=0$
and  $\Sigma_d \not=0$, with a necessary tuning of their alignment in the $U(3)_d$ space in order to avoid large 
right-handed mixing (which is strongly constrained by data). 
\end{itemize}
In view of these considerations, we consider the $U(2)^5$ case more interesting, from a model-building point of view,
with respect to a framework with some unbroken  $U(3)$ subgroups. The only variation that is worth to consider, 
in order to treat $y_b$ and/or $y_\tau$  as small parameters, is the addition of appropriate $U(1)$ groups
acting only on the right-handed $b$ and $\tau$ fields, that is what we discuss next.

\subsection{$U(2)^5 \otimes U(1)_b \otimes  U(1)_\tau$}
\label{sect:U2U1U1}

Enlarging the $U(2)^5$ symmetry with two $U(1)$ groups under which only $b_R$ and $\tau_R$ are charged,  
the only mass term allowed in the limit of unbroken symmetry is  the top Yukawa coupling. 
In order to allow non-vanishing bottom and tau Yukawa couplings, two 
spurions $X_{b(\tau)}$, breaking the $U(1)_{b(\tau)}$ groups, must be included. 
Considering these spurions, in addition to those of the minimal setup, the Yukawa matrices 
assume the form 
\begin{align}
\begin{aligned}
Y_{u}&=y_t
\begin{pmatrix}
\Delta_u & x_{t}\,V_q \\
0 & 1
\end{pmatrix}
\,,&
Y_d&= 
\begin{pmatrix}
\Delta^\prime_d &  x^\prime_b\,V_q \\
0 & \kappa_b X_b
\end{pmatrix}
\,,&
Y_e&= 
\begin{pmatrix}
\Delta^\prime_e & x^\prime_\tau\, V_\ell \\
0 & \kappa_\tau X_\tau
\end{pmatrix}
\,.
\end{aligned}
\label{eq:U2Yukawas}
\end{align} 
where $\Delta^\prime_{b(\tau)}$ differ by $\Delta_{b(\tau)}$ only by the overall normalization, 
and $\kappa_{b,\tau}$ are two additional free parameters that we are allowed to include 
in this framework, depending on the normalization of the $U(1)$-breaking terms. 
The counting of independent coefficients of SMEFT operators at each order in the spurion expansion in this setup, 
is reported in Table~\ref{tab: final results U(2)xU(1)xU(1)}, while the detailed counting order by order,
organised according to the different sub-categories of operators is presented in  Table~\ref{tab:U2U1detailed}
in Appendix~\ref{app:tables}.  As can be expected, in absence of $X$ spurions, in this framework one 
finds less terms than at corresponding order in the $V$ and  $\Delta$ expansion in the minimal $U(2)^5$ setup.
On the other hand, when going to $\cO(X^2)$ one recovers the  same numbers of independent terms 
of the minimal $U(2)^5$ setup at the corresponding order in the $V$ and  $\Delta$ expansion.

\begin{table}[t]
	\resizebox{\textwidth}{!}{
	\renewcommand{\arraystretch}{1.2} 
	\begin{tabular}{c || cc | cc | cc | cc | cc | cc | cc | cc}
		&
	\multicolumn{16}{c}{ $U(2)^5 \otimes U(1)_b \otimes U(1)_\tau$ [terms summed up to different orders]} \\[.3cm]
		& &
		& &
		& &
		& &
		& &
		& \multicolumn{2}{c|}{$\mathcal{O}{(V^2,\Delta^1,}$}
		& \multicolumn{2}{c|}{$\mathcal{O}{(\Delta^1 V^1, }$}
		& \multicolumn{2}{c}{$\mathcal{O}{(V^3,\Delta^1 V^1,}$}
		\\ 
		Operators
		& \multicolumn{2}{c|}{Exact}
		& \multicolumn{2}{c|}{$\mathcal{O}{(X^2)}$}
		& \multicolumn{2}{c|}{$\mathcal{O}{(V^1, X^2)}$}
		& \multicolumn{2}{c|}{$\mathcal{O}{(V^2,V^1 X^2)}$}
		& \multicolumn{2}{c|}{$\mathcal{O}{(\Delta^1,V^1 X^2)}$}
		& \multicolumn{2}{c|}{$\quad V^1 X^2)$}
		& \multicolumn{2}{c|}{$V^2 X^2,  \Delta^1 X^1)$}
		& \multicolumn{2}{c}{$V^2 X^2, \Delta^1 X^1) \!\!\!\! $}
		\\ \hline

		Class 1--4
		& 9 & 6 	& 9 & 6	 	& 9 & 6	 	& 9 & 6	 	& 9 & 6	 	& 9 & 6	 	& 9 & 6	 	& 9 & 6 	\\ \hline
		$\psi^2 H^3$
		& 1 & 1 	& 3 & 3 	& 4 & 4	 	& 6 & 6	 	& 9 & 9 	& 9 & 9 	& 12 & 12 	& 12 & 12 	\\
		$\psi^2 X H$
		& 3 & 3 	& 8 & 8 	& 11 & 11 	& 16 & 16 	& 24 & 24 	& 24 & 24 	& 32 & 32 	& 32 & 32 	\\
		$\psi^2 H^2 D$
		& 14 & -- 	& 15 & 1 	& 19 & 5 	& 23 & 5 	& 19 & 5 	& 23 & 5 	& 25 & 7 	& 25 & 7 	\\ \hline
		$(\bar{L}L)(\bar{L}L)$
		& 23 & -- 	& 23 & -- 	& 40 & 17 	& 67 & 24 	& 40 & 17 	& 67 & 24 	& 67 & 24 	& 74 & 31 	\\
		$(\bar{R}R)(\bar{R}R)$
		& 29 & -- 	& 29 & -- 	& 29 & -- 	& 29 & -- 	& 29 & -- 	& 29 & -- 	& 38 & 9 	& 38 & 9 	\\
		$(\bar{L}L)(\bar{R}R)$
		& 32 & -- 	& 32 & -- 	& 48 & 16 	& 64 & 16 	& 50 & 18 	& 66 & 18 	& 77 & 29 	& 77 & 29 	\\
		$(\bar{L}R)(\bar{R}L)$
		& -- & -- 	& 1 & 1	 	& 1 & 1	 	& 3 & 3 	& 3 & 3 	& 3 & 3 	& 6 & 6 	& 6 & 6 	\\
		$(\bar{L}R)(\bar{L}R)$
		& -- & -- 	& 4 & 4	 	& 4 & 4	 	& 12 & 12 	& 18 & 18 	& 18 & 18 	& 38 & 38 	& 38 & 38 	\\ \hline\hline
{\bf total:}
		& 111 & 10 	& 124 & 23 	& 165 & 64 	& 229 & 88 	& 201 & 100 & 248 & 107	& 304 & 163	& 311 & 170 	\\
	\end{tabular}
	}
	\caption{Number of independent operators at different orders in the spurion expansion for the SMEFT with $U(2)^5 \otimes U(1)^2$ flavour symmetry.
	\label{tab: final results U(2)xU(1)xU(1)}}
\end{table}

\subsection{Classification of  all possible $U(2)^5$ breaking terms.}
To conclude this section, we present a list of the possible breaking terms of $U(2)^5$ beyond the minimal setup.
The $U(2)^5$ breaking spurions can be easily classified by analysing how the bilinears $\bar\psi\psi$ and the four-fermion operators  $\bar\psi\psi\bar\psi\psi$ transform. 
Focusing on the bilinears only,\footnote{~Irreducible breaking terms of higher rank are difficult to realise in explicit models} $20$ different structure arise:
\begin{itemize}
\item 5 spurions $V_{\psi} \sim (\bar \psi \psi_3) \sim 2_\psi $, where $\psi=\{L,Q,E,U,D\}$ and $\psi_3=\{\ell_3,q_3,\tau_R,t_R,b_R\}$
(in the minimal setup only the two $V_{\psi}$ breaking the two left-handed subgroups are included);
\item 10 bi-fundamental spurions $\Delta_{\psi\psi^\prime} \sim   (\bar \psi\psi^\prime) \sim \bar 2_\psi \times 2_{\psi^\prime}$, with $\psi\ne\psi^\prime$, which include 
	6 leptoquark spurions, 3 di-quark  spurions, and 1 di-lepton spurion 
(in the minimal setup only the three Yukawa-like $\Delta_{\psi\psi^\prime}$ are included);	
\item 5 adjoints spurions $A_{\psi}\sim (\bar \psi\psi) \sim 3_\psi$ (none of which is included in the minimal setup).
\end{itemize}
Analysing the number of independent operators with all these spurions is quite straightforward using the results presented in Section~\ref{sec:U2}. However, it is less obvious how to define a consistent 
power-counting. Indeed  in the minimal setup the size of the spurions can be directly inferred by the structure of the Yukawa couplings 
(the only exception being $V_\ell$, whose size is deduced  by imposing $|V_\ell |\sim |V_q|$). Once this is fixed, the minimal setup ensures 
a CKM-like suppression for all left-handed transitions in the quark sector,
as well as the helicity suppression of mixing in the right-handed sector. On the other hand, these two important phenomenological properties 
are lost if any of the non-minimal spurions listed above can compete, in size, with the two leading breaking terms (without a special alignment
in flavour space). It is therefore natural to conclude that if any of the non-minimal spurions are included, they must be quite small in size.

\section{A phenomenological application: LFV at the LHC}
\label{sect:pheno}
The usefulness of specific hypotheses about symmetry and symmetry-breaking in the flavour sector is quite clear when analysing  
low-energy flavour-violating observables, both in the quark and in the lepton sector. However, it is worth to stress that the flavour symmetry is quite 
useful also to simplify and organise analyses of high-$p_T$ observables at the LHC.
As clearly demonstrated in~\cite{Greljo:2017vvb} with the analysis of $pp\to \mu \bar \mu$ data, 
flavour hypothesis provides a very useful organising principle to sum, with a proper weight, 
the contributions of different quarks species in $pp$ collisions.  To illustrate this statement in the context of the general formalism 
we have introduced for the $U(2)^5$ SMEFT basis, in the following we briefly discuss how to extract bounds 
on semi-leptonic four-fermion operators from the Lepton Flavour Violating (LFV) Drell-Yan process $pp\to  \tau \bar \ell $ ($\ell=e,\mu$).

In presence of $d=6$ SMEFT operators, the process $pp\to\tau \bar \ell $  receives tree-level contributions by the partonic scattering $q_i \bar q_j \to \tau \bar \ell $. 
The quark flavours $\{i, j\}$ appear in the cross-section with a weight that depends on: i) the operators we are considering,   ii)~the parton distribution functions (PDF) 
of the colliding protons. 
Hence, the physical cross-section can be written as the trace over the contraction of two flavour tensors: a ``SMEFT tensor'', that 
we denote $F^{\ell\tau}_q (\{C_i\})$  and that depends on the SMEFT coefficients, and a``PDF tensor'', that we denote $K_q$. In terms of these 
two tensor we can write
\begin{align}\label{eq:xsec}
\sigma(pp\to \tau \bar \ell )=  \frac{s}{144\pi\, \Lambda^4}\,  \mathrm{Tr}\left( F_q^{\ell \tau} (\{C_i\}) \cdot K_q  \right) \,,
\end{align}
where the trace runs over all possible pairs of colliding quarks, as we discuss below, 
$\Lambda$ is an overall scale that we introduce to normalise the coefficients of the $d=6$ operators, 
$\sqrt{s}$ is the proton-proton center-of-mass energy, and summation over the index $q=\{u,d\}$ labelling up and down quarks is implied.

\begin{itemize}
\item{} 
The SMEFT tensor $F^{\ell\tau}_q$ is in direct correspondence with the tensor $\Sigma_{\ell q}$ 
defined  in Eq.~\eqref{eq:SigmaTens}, controlling the flavour structure of the $Q_{\ell q}^{(1,3)}$
operators, and similar tensors for the other semileptonic operators in Table~\ref{tab:OperatorClasses}.
Since our scope is merely illustrative, in the following we will limit ourself to consider only 
the contributions of $Q_{\ell q}^{(1,3)}$, neglecting terms with different helicity.\footnote{In the limit of massless 
fermions the helicity is conserved and the contributions of the other semilpetonic operators 
do not interfere with that of $Q_{\ell q}^{(1,3)}$ in the cross section. }
Separating up and and down quark components of $F^{\ell\tau}_q$, as well as $SU(2)_L$ singlet and triplet components of $\Sigma_{\ell q}$,
we can write
\begin{align}
F^{\ell\tau nm}_u &=\  \Big|V^{nr}_{\text{CKM}}V^{ms\,\ast}_{\text{CKM}} \,\left( \Sigma_{\ell q}^{(1)\,\ell\tau, rs}  
-\Sigma_{\ell q}^{(3)\,\ell\tau, rs}    \right) \Big|^2\,, \no \\
F^{\ell\tau nm}_d &=\  \Big|\Sigma_{\ell q}^{(1)\,\ell\tau, nm}+\Sigma_{\ell q}^{(3)\,\ell\tau, nm}\Big|^2\,,
\label{eq:Ftensor}
\end{align} 
where summation over repeated indices is implicit. 
The $\Sigma^{(i)}_{\ell q}$ appearing in (\ref{eq:Ftensor}) are written in the down-quark mass basis, whereas the 
explicit expression displayed in Table~\ref{tab:SigmaQL} is in the interaction basis. 
After rotating to the down-quark mass-eigenstate basis (see appendix \ref{sect:appA}), the  entries of $\Sigma^{(i)}_{\ell q}$  relevant to 
 $pp\to\tau \bar \mu $ read:
\begin{align}
\Sigma_{\ell q}^{(i)\, \mu\tau, dd}\ &=\ \ \ \epsilon_\ell e^{i\bar\phi_\ell}\, c_e\, \left\{ C_{\ell q}^{(i)}(\beta_1) -s_b \epsilon_q\, s_d^2\left[ {\tilde C}_{\ell q}^{(i)}(\gamma_1) + 
{\tilde C}_{\ell q}^{(i)}(\gamma_2)  \right] + s_b^2\, s_d^2\, C_{\ell q}^{(i)}(\beta_2) \right\} \no\\
\Sigma_{\ell q}^{(i)\, \mu\tau, ds}\ &=\ -\epsilon_\ell e^{i\bar\phi_\ell}\, c_e  \,\left\{  s_b \epsilon_q\, \left[{\tilde C}_{\ell q}^{(i)}(\gamma_1)  + {\tilde C}_{\ell q}^{(i)}(\gamma_2) \right]-s_b^2\,C_{\ell q}^{(i)}(\beta_2)  \right\}\, c_d s_d\, e^{i \alpha_d} \no\\
\Sigma_{\ell q}^{(i)\, \mu\tau, db}\ &=\ \ \ \epsilon_\ell e^{i\bar\phi_\ell}\, c_e\left\{ s_b \left[C_{\ell q}^{(i)}(\beta_1)-C_{\ell q}^{(i)}(\beta_2)\right]+\epsilon_q {\tilde C}_{\ell q}^{(i)}(\gamma_1) -s_b^2\epsilon_q \,{\tilde C}_{\ell q}^{(i)}(\gamma_2)  \right\}\,s_d\, e^{i\alpha_d}  \no \\
\Sigma_{\ell q}^{(i)\, \mu\tau, sd}\ &=\ -\epsilon_\ell e^{i\bar\phi_\ell}\, c_e \,\left\{  s_b \epsilon_q\, \left[{\tilde C}_{\ell q}^{(i)}(\gamma_1)  + {\tilde C}_{\ell q}^{(i)}(\gamma_2) \right]-s_b^2\,C_{\ell q}^{(i)}(\beta_2)  \right\} \,c_d s_d\, e^{-i \alpha_d} \no\\
\Sigma_{\ell q}^{(i)\, \mu\tau, ss}\ &=\ \ \ \epsilon_\ell  e^{i\bar\phi_\ell}\,c_e\, \left\{ C_{\ell q}^{(i)}(\beta_1)-s_b \epsilon_q\, c_d^2\left[{\tilde C}_{\ell q}^{(i)}(\gamma_1)  + {\tilde C}_{\ell q}^{(i)}(\gamma_2) \right] + s_b^2\, c_d^2\,C_{\ell q}^{(i)}(\beta_2)\right\} 
\label{eq:Sigmadbase} \\
\Sigma_{\ell q}^{(i)\, \mu\tau, sb}\ &=\ \ \ \epsilon_\ell e^{i\bar\phi_\ell}\,c_e\ \left\{ s_b \left[C_{\ell q}^{(i)}(\beta_1)-C_{\ell q}^{(i)}(\beta_2)\right]+\epsilon_q {\tilde C}_{\ell q}^{(i)}(\gamma_1) -s_b^2\epsilon_q\, {\tilde C}_{\ell q}^{(i)}(\gamma_2) \right\}\, c_d  \no\\
\Sigma_{\ell q}^{(i)\, \mu\tau, bd}\ &=\ \ \ \epsilon_\ell e^{i\bar\phi_\ell}\, c_e \,\left\{ s_b \left[C_{\ell q}^{(i)}(\beta_1)-C_{\ell q}^{(i)}(\beta_2)\right]+\epsilon_q {\tilde C}_{\ell q}^{(i)}(\gamma_2) -s_b^2\epsilon_q\, {\tilde C}_{\ell q}^{(i)}(\gamma_1)  \right\}\,s_d\, e^{-i\alpha_d} \no\\
\Sigma_{\ell q}^{(i)\, \mu\tau, bs}\ &=\ \ \ \epsilon_\ell e^{i\bar\phi_\ell}\,c_e\, \left\{s_b \left[C_{\ell q}^{(i)}(\beta_1)-C_{\ell q}^{(i)}(\beta_2)\right]+\epsilon_q {\tilde C}_{\ell q}^{(i)}(\gamma_2) -s_b^2\epsilon_q \, {\tilde C}_{\ell q}^{(i)}(\gamma_1) \right\}\,c_d \no\\
\Sigma_{\ell q}^{(i)\, \mu\tau, bb}\ &=\ \ \ \epsilon_\ell e^{i\bar\phi_\ell}\, c_e\,\left\{C_{\ell q}^{(i)}(\beta_2)+s_b \epsilon_q\,\left[{\tilde C}_{\ell q}^{(i)}(\gamma_1)  + {\tilde C}_{\ell q}^{(i)}(\gamma_2) \right] + s_b^2\,C_{\ell q}^{(i)}(\beta_1)\right\}\,,  \no
\end{align}
while $\Sigma_{\ell q}^{(i)\,  e\tau, nm}=(s_e/c_e) \times \Sigma_{\ell q}^{(i)\, \mu\tau, nm}$ for $pp\to \tau \bar e$. Here 
${\tilde C}_{\ell q}^{(i)}(\gamma_{1,2})=C_{\ell q}^{(i)}(\gamma_2) e^{\pm i(\bar \phi_q -  \phi_q)}$, 
and the mixing and phase parameters $c_e$, $c_b$, $c_d$, $\phi_q$ and $\alpha_d$ are defined in appendix \ref{sect:appA}. 
To simply the expressions we have set $s_\tau=0$ and we have neglected terms proportional to 
${\tilde C}_{\ell q}^{(i)}(\xi)$, since they always appear suppressed by $\epsilon_\ell\epsilon_q^2$.

\item{}  The PDF tensor $K_q$ is given by 
\begin{align}\label{eq:lumis}
K_q^{mn}=\int\mathrm{d}\tau\, \tau\, \mathcal{L}_{q_m\bar q_n}(\tau)\,,
\end{align}
where $\tau\equiv m_{\ell\tau}^2/s$, $m_{\ell\tau}$ being the invariant mass of the lepton pair, and $\mathcal{L}_{q_m\bar q_n}$ are the parton luminosity functions for ${\bar q_n q_m}$ colliding partons defined by 
 \begin{align}
\mathcal{L}_{q_m\bar q_n}(\tau)=\int_\tau^1\frac{\mathrm{d}x}{x}\, \left[\,f_{q_m}\!(x,\mu_F)\,f_{\bar q_{n}}\!\left(\tau/x,\mu_F\right)\ +\ (q_n\!\leftrightarrow\! \bar q_m)\,\right]\, ,
\end{align}
in terms of the single parton PDF $f^a$ ($\mu_F$ denotes the factorization scale). 
Again for illustrative purposes,  we compute \eqref{eq:lumis} by integrating the high-mass tail of $m_{\ell\tau}$ in the range $m_{\ell\tau}\in[1,5]$~TeV 
at $\sqrt{s}=13$~TeV, setting $\mu_F= s\tau$ and using the central values of the {\tt PDF4LHC15$\,\!\_\,$nnlo$\,\!\_\,$mc} PDF set \cite{Butterworth:2015oua}
This leads to
\begin{align}\label{eq:PDFflavor}
K_u \ =\ \kappa\,\left(\begin{array}{ccc} 1& 0.5 &0 \\ 0.03& 0.01& 0 \\ 0 & 0 & 0\end{array}\right)\ \ , \ \ \ \ \ \ \
K_d \ =\ \frac{\kappa}{2}\,\left(\begin{array}{ccc} 1&0.6&0.3\\0.1&0.07&0.03\\0.04&0.02&0.01\end{array}\right)\,, 
\end{align}
with $\kappa\approx 4.8\times 10^{-3}$. The third row and column for $K_u$ vanishes because the PDF of the top-quark in  still 
negligible at  $\sqrt{s}=13$~TeV.
\end{itemize}

We have now all the ingredients to compute the traces in \eqref{eq:xsec}. In the down sector, the explicit calculation yields
\begin{eqnarray}
&& \mathrm{Tr}\left( F_d^{\ell \tau}\cdot K_d  \right)
= \sum_{n,m=\{d,s,b\}}\Big|\Sigma^{\ell\tau,nm}_{\ell q}\Big|^2\,K^{mn}  \no \\
\no\\
&&\qquad  \propto\  \left(1.07+0.068\, s_b^2 \right)\, \left| C_{\ell q}^{(1+3)}(\beta_1) \right|^2  + \left(0.01+0.068\, s_b^2 \right)\, \left| C_{\ell q}^{(1+3)}(\beta_2) \right|^2
\no \\
&&\qquad\quad  + 0.13 \,s_b^2~{\rm Re}\left[ C_{\ell q}^{(1+3)}(\beta_1) C_{\ell q}^{(1+3)}(\beta_2)^* \right] 
 - s_b \epsilon_q ~{\rm Re}\Big[ 0.15~C_{\ell q}^{(1+3)}(\beta_1) {\tilde C}_{\ell q}^{(1+3)}(\gamma_2)^* \no \\
 &&\qquad\quad
 + 0.2 ~C_{\ell q}^{(1+3)}(\beta_1) {\tilde C}_{\ell q}^{(1+3)}(\gamma_1) ^*+ 0.072 ~C_{\ell q}^{(1+3)}(\beta_2) {\tilde C}_{\ell q}^{(1+3)}(\gamma_2) ^*+ 0.024 C_{\ell q}^{(1+3)}(\beta_2) {\tilde C}_{\ell q}^{(1+3)}(\gamma_1) ^* \Big]
\no\\
&&\qquad\quad  + \epsilon_q^2~ \left[  0.046~ \left| {\tilde C}_{\ell q}^{(1+3)}(\gamma_2) \right|^2 +  0.022~ \left|{\tilde C}_{\ell q}^{(1+3)}(\gamma_1) \right|^2 \right]
+\cO(\epsilon^3_q) \\
\nonumber\\
&&\qquad \stackrel{\gamma_i =0}{\longrightarrow}  \quad 
\left(1.07+0.068\, s_b^2 \right)\, \left|C_{\ell q}^{(1+3)}(\beta_1) \right|^2  + \left(0.01+0.068\, s_b^2 \right)\, \left| C_{\ell q}^{(1+3)}(\beta_2) \right|^2 \no\\
&&\qquad\qquad\quad + 0.13 \,s_b^2~{\rm Re}\left[ C_{\ell q}^{(1+3)}(\beta_1) C_{\ell q}^{(1+3)}(\beta_2)^* \right]~, \qquad
\end{eqnarray}
where we used $s_d\approx|V_{td}|/|V_{ts}|$ and $c_d\approx1$ (see appendix \ref{sect:appA}), we have factored out the leptonic part,
common to all operators, and we have taken into account that $s_b= \cO(\epsilon_q)$ when neglecting $\cO(\epsilon^3_q)$ terms.  
Here and below $C_{\ell q}^{(1\pm 3)}(F) = C_{\ell q}^{(1)}(F) \pm C_{\ell q}^{(3)}(F)$. 
Similarly, from the up-quarks we get 
\begin{eqnarray}
&& \mathrm{Tr}\left( F_u^{\ell \tau}\cdot K_u  \right)
= \sum_{n,m=\{u,c\}}\Big|V_{\text{CKM}}^{nr}V_{\text{CKM}}^{ms\,*}\,\Sigma^{\ell\tau,rs}_{\ell q}\Big|^2\,K^{mn} \no\\
\no\\
&&\qquad  \propto\   1.01~ \left| C_{\ell q}^{(1-3)}(\beta_1) \right|^2  - (0.003~s_b -0.035~ s_b^2)~{\rm Re}\left[ C_{\ell q}^{(1-3)}(\beta_1) C_{\ell q}^{(1-3)}(\beta_2)^*
\right] \\
 &&\qquad\quad   - 0.036~s_b \epsilon_q ~{\rm Re}\left[ C_{\ell q}^{(1-3)}(\beta_1) {\tilde C}_{\ell q}^{(1-3)}(\gamma_2) ^* + C_{\ell q}^{(1-3)}(\beta_1) {\tilde C}_{\ell q}^{(1-3)}(\gamma_1)^* \right] + \cO(\epsilon^3_q)~,
 \end{eqnarray}
where we have implemented the numerical values of the CKM matrix from~\cite{Tanabashi:2018oca} and, consistently with the assumption of neglecting 
$\cO(\epsilon^3_q)$ terms, we have neglected numerical entries of $\cO(10^{-3})$.

From the above expressions we can draw the following conclusions:
\begin{itemize}
\item{}  Operators which are both quark- and lepton-flavour violating (parameterised by the coefficients $\gamma_i$) 
contribute with very small numerical coefficients (recall that $|\epsilon_q|, |s_b| \lsim  10^{-1} $).
Once we take into account the corresponding bounds from low-energy processes, such as $B_s \to \tau \ell$, their impact in the cross-section is negligible.
This is a nice virtue of the $U(2)^5$ approach that, contrary to the MFV case, allow us to easily separate flavour-conserving and flavour-violating operators.
\item{} Once quark-violating terms are eliminated, we remain with an expression depending only on 
 $C_{\ell q}^{(i)}(\beta_1) $ and  $C_{\ell q}^{(i)}(\beta_2) $, which is 
quite simple. Here the contribution of  $C_{\ell q}^{(i)}(\beta_1)$ is largely dominant receiving contributions by valence 
quarks collisions ($u\bar u, d\bar d\to \tau \bar \ell $).\footnote{The coefficients $C_{\ell q}^{(i)}(\beta_{1,2})$ also appear in 
low-energy LFV processes such as $\tau\to \rho \ell$ and $\Upsilon \to \bar \tau \ell$. A detailed analysis of these low-energy processes 
is beyond the scope  of this work. We simply note that, at present, the low-energy bounds on $C_{\ell q}^{(1+3)}(\beta_1)$
are more stringent (by less than one order of magnitude)  than those which can be extracted from $\sigma(pp \to \tau \bar \ell )$,
 whereas collider bounds on $C_{\ell q}^{(1+3)}(\beta_2)$ dominate vs.~the low-energy ones \cite{Angelescu:2020uug}.}

\item{} The other virtue of the  $U(2)^5$ approach is that of separating the contribution of non-trivial dynamics associated to the third generation, 
which in many UV completions of the SM is the one associated to larger couplings (or lower effective scales). In our specific example these are the 
terms proportional to $|C_{\ell q}^{(i)}(\beta_2) |^2$. In this case it is interesting to 
analyse the relative weight of third-generation partons, which have suppressed PDFs, vs.~the flavour-violating terms, which receive contributions 
from the larger light-quark PDFs. The analytical expression of the terms proportional to $|C_{\ell q}^{(i)}(\beta_2) |^2$ in the cross section is 
\begin{align}\label{eq:3rdgenNP}
\left. \mathrm{Tr}\left( F_d^{\ell \tau}\cdot K_d  \right)\right|_{|C_{\ell q}^{(i)}(\beta_2) |^2}  \propto\ K_d^{bb} \ &+\ \left[K_d^{bs}+K_d^{sb}+\frac{|V_{td}|^2}{|V_{ts}|^2} 
\left(K_d^{bd}+K_d^{db}\right)\right] s_b^2\\
&+\ \left[K_d^{ss} + \frac{|V_{td}|^2}{|V_{ts}|^2} \left(K_d^{ds}+K_d^{sd}\right)+\frac{|V_{td}|^4}{|V_{ts}|^4} K_d^{dd}\right] s_b^4\,.
\end{align}
Given $|s_b| \lsim  10^{-1}$, and given the entries of $K_d$ in (\ref{eq:PDFflavor}), it is easy to realise that for this specific operator the cross-section 
is completely dominated by the $b\bar b\to \tau \bar \ell $ channel, a conclusion that is not obvious a priori and that follows from 
the underlying flavour symmetry and symmetry-breaking ansatz. 
\end{itemize}

\section{Conclusions}

In this paper we have analysed how the flavour symmetries $U(3)^5$ and $U(2)^5$, with a minimal set of breaking terms,
act as organising principle for the dimension-six operators with flavour quantum numbers in the SMEFT. 
The main results of our analysis 
are summarised in Table~\ref{tab:U3new}  and~\ref{tab: final results}. As can be seen from Table~\ref{tab:U3new}, the $U(3)^5$ symmetry, and the 
MFV hypothesis, allow us to obtain a drastic reduction in the number of independent terms: this is about 25 times smaller compared to the case 
of no flavour symmetry (limiting ourself to the leading terms in the MFV expansion).
In the $U(2)^5$ case, which encompass a wider spectrum of new-physics models where the third generation plays a special role, 
the number of independent terms is roughly three times higher compared to the MFV case (see Table~\ref{tab: final results}),  
but it is still about one order of magnitude smaller 
compared to the case of no flavour symmetry.

We have also provided a general discussion of why these two sets of flavour hypotheses (i.e.~flavour symmetry and symmetry-breaking ansatz)
are particularly relevant if we are interested in SMEFT implementations where the bounds on the scale of new physics are 
not largely saturated by the tight constraints from flavour-violating processes. The two flavour hypotheses 
we have considered in detail are somehow unique in ensuring a ``natural'' CKM-like suppression 
for all left-handed transitions in the quark sector, as well as a double (light-mass~$\times$~CKM) suppression for mixing 
in the right-handed quark sector. Variations preserving these properties are possible with the implementation of  $U(1)$ symmetries 
acting on the third family, as in the case of the $U(2)^5 \times U(1)_b\times U(1)_\tau$ group
discussed in some detail in Section~\ref{sect:beyondU2}. In principle, more options are possible in the lepton sector, where we have not made any attempt to 
link flavour violation in the charged-lepton sector to neutrino mixing, which goes beyond the scope of the present analysis.

To highlight the usefulness of this approach  in analysing   high-$p_T$ observables, and the practical implementation 
of the classification scheme we have introduced for the SMEFT with minimally broken $U(2)^5$,
we have discussed the concrete example of $\sigma(pp\to \tau \bar \ell )$.
As already shown in~\cite{Greljo:2017vvb} for $\sigma(pp\to \mu \bar \mu)$,
also for $\sigma(pp\to \tau \bar \ell )$  
the implementation of flavour symmetries  provides an essential tool to sum
the contributions of the different quarks species appearing in $pp$ collisions,  
which a priori would involve a large number of independent SMEFT coefficients.
This is quite relevant given the growing impact of high-$p_T$ data in constraining four-fermion operators 
and, particularly, the flavour-violating ones~\cite{Angelescu:2020uug,Fuentes-Martin:2020lea}.

So far, most of the implementations of $U(3)^5$ and $U(2)^5$ symmetries in the SMEFT, or better in 
subsets of the SMEFT operators, were aimed at analysing low-energy data only. However, the key virtue 
of both these approach is that of generating an EFT where bounds from flavour-violating and 
flavour-conserving observables are naturally very complementary,
as shown for instance in~\cite{Greljo:2017vvb,Aoude:2020dwv,Angelescu:2020uug}. Most important, with the 
implementations of these flavour hypotheses the overall  effective scale of the SMEFT can be 
kept relatively low, making the whole construction quite interesting for phenomenology.
  We thus believe the classification we have presented 
in this paper can be a useful first step toward a systematic analysis
of both flavour-conserving and flavour-violating observables, at both low- and high-energies, 
in motivated flavoured versions of the SMEFT.

\subsubsection*{Acknowledgments}

We thank Riccardo Barbieri,  Ilaria Brivio,  Admir Greljo, Marco Nardecchia, and Sophie Renner, for useful discussion at different stages of this work.
This project has received funding from the  European Research Council (ERC) under the European Union's Horizon 2020 research and innovation programme  under grant agreement 833280 (FLAY), and by the Swiss National Science Foundation (SNF) under contract 200021-159720. 
The work of K.Y. was also supported in part by the JSPS KAKENHI 18J01459.

\bibliographystyle{JHEP}

{\footnotesize
\bibliography{U2SMEFT_v9}

\providecommand{\href}[2]{#2}\begingroup\raggedright\begin{thebibliography}{10}

\bibitem{Buchmuller:1985jz}
W.~Buchmuller and D.~Wyler, \emph{{Effective Lagrangian Analysis of New
  Interactions and Flavor Conservation}},
  \href{http://dx.doi.org/10.1016/0550-3213(86)90262-2}{\emph{Nucl. Phys.} {\bf
  B268} (1986) 621--653}.

\bibitem{Grzadkowski:2010es}
B.~Grzadkowski, M.~Iskrzynski, M.~Misiak and J.~Rosiek, \emph{{Dimension-Six
  Terms in the Standard Model Lagrangian}},
  \href{http://dx.doi.org/10.1007/JHEP10(2010)085}{\emph{JHEP} {\bf 10} (2010)
  085}, [\href{http://arxiv.org/abs/1008.4884}{{\tt 1008.4884}}].

\bibitem{Jenkins:2013zja}
E.~E. Jenkins, A.~V. Manohar and M.~Trott, \emph{{Renormalization Group
  Evolution of the Standard Model Dimension Six Operators I: Formalism and
  lambda Dependence}},
  \href{http://dx.doi.org/10.1007/JHEP10(2013)087}{\emph{JHEP} {\bf 10} (2013)
  087}, [\href{http://arxiv.org/abs/1308.2627}{{\tt 1308.2627}}].

\bibitem{Jenkins:2013wua}
E.~E. Jenkins, A.~V. Manohar and M.~Trott, \emph{{Renormalization Group
  Evolution of the Standard Model Dimension Six Operators II: Yukawa
  Dependence}}, \href{http://dx.doi.org/10.1007/JHEP01(2014)035}{\emph{JHEP}
  {\bf 01} (2014) 035}, [\href{http://arxiv.org/abs/1310.4838}{{\tt
  1310.4838}}].

\bibitem{Alonso:2013hga}
R.~Alonso, E.~E. Jenkins, A.~V. Manohar and M.~Trott, \emph{{Renormalization
  Group Evolution of the Standard Model Dimension Six Operators III: Gauge
  Coupling Dependence and Phenomenology}},
  \href{http://dx.doi.org/10.1007/JHEP04(2014)159}{\emph{JHEP} {\bf 04} (2014)
  159}, [\href{http://arxiv.org/abs/1312.2014}{{\tt 1312.2014}}].

\bibitem{Brivio:2017vri}
I.~Brivio and M.~Trott, \emph{{The Standard Model as an Effective Field
  Theory}}, \href{http://dx.doi.org/10.1016/j.physrep.2018.11.002}{\emph{Phys.
  Rept.} {\bf 793} (2019) 1--98}, [\href{http://arxiv.org/abs/1706.08945}{{\tt
  1706.08945}}].

\bibitem{Falkowski:2017pss}
A.~Falkowski, M.~Gonzalez-Alonso and K.~Mimouni, \emph{{Compilation of
  low-energy constraints on 4-fermion operators in the SMEFT}},
  \href{http://dx.doi.org/10.1007/JHEP08(2017)123}{\emph{JHEP} {\bf 08} (2017)
  123}, [\href{http://arxiv.org/abs/1706.03783}{{\tt 1706.03783}}].

\bibitem{Descotes-Genon:2018foz}
S.~Descotes-Genon, A.~Falkowski, M.~Fedele, M.~Gonzalez-Alonso and J.~Virto,
  \emph{{The CKM parameters in the SMEFT}},
  \href{http://dx.doi.org/10.1007/JHEP05(2019)172}{\emph{JHEP} {\bf 05} (2019)
  172}, [\href{http://arxiv.org/abs/1812.08163}{{\tt 1812.08163}}].

\bibitem{Falkowski:2019hvp}
A.~Falkowski and D.~Straub, \emph{{Flavourful SMEFT likelihood for Higgs and
  electroweak data}},  \href{http://arxiv.org/abs/1911.07866}{{\tt
  1911.07866}}.

\bibitem{Aoude:2020dwv}
R.~Aoude, T.~Hurth, S.~Renner and W.~Shepherd, \emph{{The impact of flavour
  data on global fits of the MFV SMEFT}},
  \href{http://arxiv.org/abs/2003.05432}{{\tt 2003.05432}}.

\bibitem{Isidori:2010kg}
G.~Isidori, Y.~Nir and G.~Perez, \emph{{Flavor Physics Constraints for Physics
  Beyond the Standard Model}},
  \href{http://dx.doi.org/10.1146/annurev.nucl.012809.104534}{\emph{Ann. Rev.
  Nucl. Part. Sci.} {\bf 60} (2010) 355},
  [\href{http://arxiv.org/abs/1002.0900}{{\tt 1002.0900}}].

\bibitem{Chivukula:1987py}
R.~Chivukula and H.~Georgi, \emph{{Composite Technicolor Standard Model}},
  \href{http://dx.doi.org/10.1016/0370-2693(87)90713-1}{\emph{Phys.\ Lett.\ B}
  {\bf 188} (1987) 99--104}.

\bibitem{DAmbrosio:2002vsn}
G.~D'Ambrosio, G.~F. Giudice, G.~Isidori and A.~Strumia, \emph{{Minimal flavor
  violation: An Effective field theory approach}},
  \href{http://dx.doi.org/10.1016/S0550-3213(02)00836-2}{\emph{Nucl. Phys.}
  {\bf B645} (2002) 155--187}, [\href{http://arxiv.org/abs/hep-ph/0207036}{{\tt
  hep-ph/0207036}}].

\bibitem{Barbieri:2011ci}
R.~Barbieri, G.~Isidori, J.~Jones-Perez, P.~Lodone and D.~M. Straub,
  \emph{{$U(2)$ and Minimal Flavour Violation in Supersymmetry}},
  \href{http://dx.doi.org/10.1140/epjc/s10052-011-1725-z}{\emph{Eur. Phys. J.}
  {\bf C71} (2011) 1725}, [\href{http://arxiv.org/abs/1105.2296}{{\tt
  1105.2296}}].

\bibitem{Barbieri:2012uh}
R.~Barbieri, D.~Buttazzo, F.~Sala and D.~M. Straub, \emph{{Flavour physics from
  an approximate $U(2)^3$ symmetry}},
  \href{http://dx.doi.org/10.1007/JHEP07(2012)181}{\emph{JHEP} {\bf 07} (2012)
  181}, [\href{http://arxiv.org/abs/1203.4218}{{\tt 1203.4218}}].

\bibitem{Blankenburg:2012nx}
G.~Blankenburg, G.~Isidori and J.~Jones-Perez, \emph{{Neutrino Masses and LFV
  from Minimal Breaking of $U(3)^5$ and $U(2)^5$ flavor Symmetries}},
  \href{http://dx.doi.org/10.1140/epjc/s10052-012-2126-7}{\emph{Eur.\ Phys.\
  J.\ C} {\bf 72} (2012) 2126}, [\href{http://arxiv.org/abs/1204.0688}{{\tt
  1204.0688}}].

\bibitem{Greljo:2015mma}
A.~Greljo, G.~Isidori and D.~Marzocca, \emph{{On the breaking of Lepton Flavor
  Universality in B decays}},
  \href{http://dx.doi.org/10.1007/JHEP07(2015)142}{\emph{JHEP} {\bf 07} (2015)
  142}, [\href{http://arxiv.org/abs/1506.01705}{{\tt 1506.01705}}].

\bibitem{Barbieri:2015yvd}
R.~Barbieri, G.~Isidori, A.~Pattori and F.~Senia, \emph{{Anomalies in
  $B$-decays and $U(2)$ flavour symmetry}},
  \href{http://dx.doi.org/10.1140/epjc/s10052-016-3905-3}{\emph{Eur. Phys. J.
  C} {\bf 76} (2016) 67}, [\href{http://arxiv.org/abs/1512.01560}{{\tt
  1512.01560}}].

\bibitem{Buttazzo:2017ixm}
D.~Buttazzo, A.~Greljo, G.~Isidori and D.~Marzocca, \emph{{B-physics anomalies:
  a guide to combined explanations}},
  \href{http://dx.doi.org/10.1007/JHEP11(2017)044}{\emph{JHEP} {\bf 11} (2017)
  044}, [\href{http://arxiv.org/abs/1706.07808}{{\tt 1706.07808}}].

\bibitem{Froggatt:1978nt}
C.~Froggatt and H.~B. Nielsen, \emph{{Hierarchy of Quark Masses, Cabibbo Angles
  and CP Violation}},
  \href{http://dx.doi.org/10.1016/0550-3213(79)90316-X}{\emph{Nucl. Phys. B}
  {\bf 147} (1979) 277--298}.

\bibitem{Bordone:2019uzc}
M.~Bordone, O.~Cat\`a and T.~Feldmann, \emph{{Effective Theory Approach to New
  Physics with Flavour: General Framework and a Leptoquark Example}},
  \href{http://dx.doi.org/10.1007/JHEP01(2020)067}{\emph{JHEP} {\bf 01} (2020)
  067}, [\href{http://arxiv.org/abs/1910.02641}{{\tt 1910.02641}}].

\bibitem{Brivio:2017btx}
I.~Brivio, Y.~Jiang and M.~Trott, \emph{{The SMEFTsim package, theory and
  tools}}, \href{http://dx.doi.org/10.1007/JHEP12(2017)070}{\emph{JHEP} {\bf
  12} (2017) 070}, [\href{http://arxiv.org/abs/1709.06492}{{\tt 1709.06492}}].

\bibitem{Feldmann:2008ja}
T.~Feldmann and T.~Mannel, \emph{{Large Top Mass and Non-Linear Representation
  of Flavour Symmetry}},
  \href{http://dx.doi.org/10.1103/PhysRevLett.100.171601}{\emph{Phys. Rev.
  Lett.} {\bf 100} (2008) 171601}, [\href{http://arxiv.org/abs/0801.1802}{{\tt
  0801.1802}}].

\bibitem{Kagan:2009bn}
A.~L. Kagan, G.~Perez, T.~Volansky and J.~Zupan, \emph{{General Minimal Flavor
  Violation}}, \href{http://dx.doi.org/10.1103/PhysRevD.80.076002}{\emph{Phys.
  Rev. D} {\bf 80} (2009) 076002}, [\href{http://arxiv.org/abs/0903.1794}{{\tt
  0903.1794}}].

\bibitem{Butterworth:2015oua}
J.~Butterworth et~al., \emph{{PDF4LHC recommendations for LHC Run II}},
  \href{http://dx.doi.org/10.1088/0954-3899/43/2/023001}{\emph{J. Phys. G} {\bf
  43} (2016) 023001}, [\href{http://arxiv.org/abs/1510.03865}{{\tt
  1510.03865}}].

\bibitem{Tanabashi:2018oca}
{\scshape Particle Data Group} collaboration, M.~Tanabashi et~al.,
  \emph{{Review of Particle Physics}},
  \href{http://dx.doi.org/10.1103/PhysRevD.98.030001}{\emph{Phys. Rev. D} {\bf
  98} (2018) 030001}.

\bibitem{Greljo:2017vvb}
A.~Greljo and D.~Marzocca, \emph{{High-$p_T$ dilepton tails and flavor
  physics}}, \href{http://dx.doi.org/10.1140/epjc/s10052-017-5119-8}{\emph{Eur.
  Phys. J. C} {\bf 77} (2017) 548},
  [\href{http://arxiv.org/abs/1704.09015}{{\tt 1704.09015}}].

\bibitem{Angelescu:2020uug}
A.~Angelescu, D.~A. Faroughy and O.~Sumensari, \emph{{Lepton Flavor Violation
  and Dilepton Tails at the LHC}},  \href{http://arxiv.org/abs/2002.05684}{{\tt
  2002.05684}}.

\bibitem{Fuentes-Martin:2020lea}
J.~Fuentes-Martin, A.~Greljo, J.~Martin~Camalich and J.~D. Ruiz-Alvarez,
  \emph{{Charm Physics Confronts High-$p_T$ Lepton Tails}},
  \href{http://arxiv.org/abs/2003.12421}{{\tt 2003.12421}}.

\bibitem{Fuentes-Martin:2019mun}
J.~Fuentes-Martin, G.~Isidori, J.~Pages and K.~Yamamoto, \emph{{With or without
  U(2)? Probing non-standard flavor and helicity structures in semileptonic B
  decays}},
  \href{http://dx.doi.org/10.1016/j.physletb.2019.135080}{\emph{Phys.\ Lett.\
  B} {\bf 800} (2020) 135080}, [\href{http://arxiv.org/abs/1909.02519}{{\tt
  1909.02519}}].

\end{thebibliography}\endgroup
}

\newpage
\appendix 

\section{Diagonalization of the Yukawa matrices in $U(2)^5$.}
\label{sect:appA}

Given the flavor structure of $U(2)^5$, it is convenient to parametrise unitary matrices in the following block form:
\begin{align}\label{eq:block_mix}
W=\left(\begin{array}{c|c}
B & \Theta \\[0.3em]
\hline
\\[-1em]
\Xi^\dagger& b
\end{array}\right)\,,
\end{align}
where $W$ is a generic $3\times3$ unitary matrix that we decompose into a $2\times2$ non-unitary block-matrix $B$,  $2$-dimensional column vectors $\Xi$ and $\Theta$, and a complex number $b$. Using the unitarity constraint $W^\dagger W=W W^\dagger=\mathbb{1}$ one can solve some of the blocks, in particular:
\begin{align}\label{eq:block_sol}
b=c_\theta \, e^{-i\phi} \,, \quad\quad \Xi=-\frac{e^{i\phi} }{c_\theta}\, B^\dagger\cdot\Theta
\end{align}
where $c_\theta=\cos\theta\equiv\sqrt{1-{\Theta}^\dagger\Theta}$ is an angle and $\phi$ is the unitary phase $\text{Det}(W)=e^{-i\phi}$. The block $B$ can be further decomposed into polar form as $B=(\mathbb{1}- A)\cdot U$, where $U$ is a $2\times2$ unitary matrix and $(\mathbb{1}- A)$ is a positive semi-definite hermitian matrix, satisfying $(\mathbb{1}-A)^2=\mathbb{1}-\Theta{\Theta}^\dagger$. The hermitian matrix $A$ measures the deviation of the block $B$ from being a unitary submatrix. Solving for $A$ yields:
%
\begin{align}
A\ =\  \frac{\Theta\Theta^\dagger}{1+ c_\theta}\,.
\end{align}
With this, one finds an exact expression for an arbitrary $3\times 3$  unitary matrix in block-form:
\begin{align}\label{eq:param}
W\ =\  \left(\begin{array}{c|c}
\mathbb{1}\,-\,\frac{\Theta\Theta^\dagger}{1\,+\,c_\theta} &  e^{i\phi}\,\Theta \,\\
\hline
 -e^{-i\phi}\,\Theta^\dagger& c_\theta
\end{array}\right)\cdot \left(\begin{array}{c|c}
U&\mathbb{0} \\
\hline
\mathbb{0}&  e^{-i\phi}
\end{array}\right)\,.
\end{align}
On general grounds, the block vector $\Theta$ can be parametrized with two angles $\theta$ and $\varphi$ and two complex phases:
\begin{align}\label{eq:theta}
\Theta=\,\left(\begin{array}{c} s_\theta s_\varphi\, e^{i\xi_1} \\ s_\theta c_\varphi\, e^{i\xi_2}\end{array}\right)
\end{align} 
In the limit where $\varphi\to0$ ($\varphi\to\frac{\pi}{2}$) the angle $\theta$ corresponds to a rotation in the $23$-plane ($13$-plane).

We can use this general decomposition to compute the unitary matrices that diagonalise the Yukawa couplings
starting from the interaction basis,
\be
 L_f^\dagger Y_f R_f  = {\rm diag}(y_{f_1}, y_{f_2}, y_{f_3})~, 
 \ee
for $f=d,u,e$.  The unitary matrices $L_f$ and $R_f$ are parametrized using \eqref{eq:param} with blocks $\Theta_f^{L,R}$, $U_f^{L,R}$ and $\phi^{L,R}_f$, respectively.  Following Ref.~\cite{Fuentes-Martin:2019mun}, the left handed matrices $L_f$ are to a very good approximation a product of two consecutive rotations in the $12$ and $23$ planes. The matrix $U_f^L$ is parametrized with one angle $\theta_f$ (rotation in the 12-plane) and one phase $\alpha_f$ as in \eqref{eq:Uq}. The rotation in the $23$-plane is achieved for $\varphi_f^{L,R}\!\approx\!0$ in \eqref{eq:theta}, we therefore write  $\Theta^L_d=(0,s_b)^T$, $\Theta^L_u=(0,s_t)^T$, $\Theta^L_e=(0,s_\tau)^T$ (where we set $\xi_{2\,f}^{L,R}=0$ without loss of generality). The right-handed unitary matrices $R_f$, on the other hand, are to a good approximation just a rotation in the $23$-plane, hence $U^R_f\approx\mathbb{1}$. The right-handed and left-handed angles are related through $\Theta^R_d=\text{diag}(\frac{m_d}{m_b},\frac{m_s}{m_b})\Theta^L_d$, $\Theta^R_u=\text{diag}(\frac{m_u}{m_t},\frac{m_c}{m_t})\Theta_u^L$ and $\Theta^R_e=\text{diag}(\frac{m_e}{m_\tau},\frac{m_\mu}{m_\tau})\Theta_e^L$. For quarks, the phases can be brought to satisfy $\phi_d^L=\phi_d^R=\phi_u^R\equiv \phi_q$, while all leptonic phases can be set to zero. With these settings, neglecting tiny subleading terms, we get~\cite{Fuentes-Martin:2019mun}
\begin{align}\label{eq:YukRot}
\begin{aligned}
 L_d  &\approx 
\begin{pmatrix}
 c_d   &  -s_d\,e^{i\alpha_d}  & 0  \\
 s_d\,e^{-i\alpha_d}   &  c_d &  s_b   \\
-s_d\,s_b\,e^{-i(\alpha_d+\phi_q)}   & -c_d\,s_b\, e^{-i\phi_q} & e^{-i\phi_q}
\end{pmatrix}
\,,\\[5pt]
 R_d  &\approx  
\begin{pmatrix}
1   &  0  & 0  \\
0   &  1  & \frac{m_s}{m_b}\,s_b   \\
0  & -\frac{m_s}{m_b}\,s_b\,e^{-i\phi_q} & e^{-i\phi_q}
\end{pmatrix}
\,,\\[5pt]
 R_u  &\approx
\begin{pmatrix}
1   &  0  & 0  \\
0   &  1  & \frac{m_c}{m_t}\,s_t   \\
0  & -\frac{m_c}{m_t}\,s_t\,e^{-i\phi_q} & e^{-i\phi_q}
\end{pmatrix}
\,,\\[5pt]
 L_e  &\approx    
\begin{pmatrix}
c_e       & -s_e          & 0  \\
s_e &  c_e           & s_\tau \\
-s_e s_\tau   &  -c_es_\tau & 1
\end{pmatrix}\,,\\[5pt]
 R_e  &\approx  
\begin{pmatrix}
 1   & 0 & 0  \\
0 &  1& \frac{m_\mu}{m_\tau}\, s_\tau \\
0   &  -\frac{m_\mu}{m_\tau}\, s_\tau & 1
\end{pmatrix}\,,
\end{aligned}
\end{align}
with $L_u$ constrained by $L_u=L_d\,V_{\rm CKM}^\dagger$. 
\begin{itemize}
\item{} In the quark sector the following two parameters appearing in  $L_{u,d}$ and $R_{u,d}$,
remains unconstrained:  the real parameter $s_b$ and the complex phase $\phi_q = \phi_b -\phi_t$,
defined by 
\be
\frac{s_b}{c_b} e^{i \phi_b} = x_b V_q~, \qquad \frac{s_t}{c_t} e^{ i\phi_t} = x_t V_q~, \qquad {\rm with} \qquad 
| s_t - s_b  e^{i \phi_q} |  = |V_{cb}|~,
\ee
as well ads the phase of $V_q$ (which can be chosen to make $x_t V_q$ real and positive).
On the other hand, the parameters of $U_{u(d)}$ are
completely determined via the relations
\be
\frac{s_d}{c_d} = \frac{|V_{td}|}{|V_{ts}|}~, \qquad  \frac{s_u}{c_u} = \frac{|V_{ub}|}{|V_{cb}|}~,    \qquad
\alpha_d=\arg\left( \frac{ V_{td}^*}{  V_{ts}^*} \right)~,
\qquad  \alpha_u=\arg\left( \frac{ V_{ub}}{  V_{cb}} \right)~,
\ee
which imply
\be
U_u U^\dagger_d = \left(\begin{matrix} \cos \theta_c   & \sin \theta_c   \\ -  \sin \theta_c   & \cos \theta_c  \end{matrix} \right)~
\ee
where $\sin\theta_c = |V_{us}|$.
\item{} 
In the lepton sector both $s_\tau/c_\tau  = | x_\tau V_\ell |$  and  $s_e$  
remain unconstrained, together with the phase of $V_\ell$ (which can be chosen to  make $x_\tau V_\ell$ real and positive).
\end{itemize}

\section{On the number of independent fermion contractions}
\label{app:contractions}
Since $U(N) \sim U(1) \times SU(N)$, to verify that any of the SMEFT operators we consider transforms as a singlet under the flavour group, we have to ensure two things. First, the total phase associated to the $U(1)$ part of the flavour symmetry must be zero. Second, the operator has to be a singlet of the $SU(N)$ group. 
%
%
To determine all ways to form singlets under the $SU(N)$ groups we first consider the tensor products of two (anti-)fermion fields in the $N=3$ and $N=2$ case
\begin{align}
	\text{For}\ SU(3): && 3 \otimes \bar{3} = 1 \oplus 8, && 3 \otimes 3 = \bar{3} \oplus 6, && \bar{3} \otimes \bar{3} = 3 \oplus \bar{6}, \nonumber \\
	\text{For}\ SU(2): && 2 \otimes \bar{2} = 1 \oplus 3, && 2 \otimes 2 = 1 \oplus 3, && \bar{2} \otimes \bar{2} = 1 \oplus 3.
\end{align}
For $SU(3)$ we only get a singlet by contracting a fermion with an anti-fermion, whereas for $SU(2)$ this is also obtained for two fermions or two anti-fermions 
However, additional operators that would be allowed due to that are forbidden by the phase factors, e.g~$L L$ or~$\bar{L}^c L$ have non-vanishing total phase. 
Thus, the only allowed operators bilinear in the fermion fields have each a fermion and an anti-fermion in both cases.

Given the above consideration, for the operators involving four fermion fields we have to pay special attention only to those with four fermion fields of the same type
(i.e.~which $Q_{\ell\ell},\ Q_{qq}^{(1)},\ Q_{qq}^{(3)},\ Q_{ee},\ Q_{uu}$ and~$Q_{dd}$).  In the following we will consider $Q_{\ell\ell}$ as representative example. 
In the $N=2$ case, at leading order in the spurions, we have different choices for the contraction of the flavour indices
\begin{align}
	(i):\quad \left(\bar{L}^{r}L^{r}\right) \left(\bar{L}^{s}L^{s}\right), &&
	(ii):\quad \left(\bar{L}^{r}L^{s}\right) \left(\bar{L}^{s}L^{r}\right), &&
	(iii):\quad \epsilon^{r_1\, r_2}\epsilon^{s_1\, s_2} \left(\bar{L}^{r_1}L^{s_1}\right) \left(\bar{L}^{r_2}L^{s_2}\right),
	\label{eq: U(2) quadrilinear contractions}
\end{align}
where the letter corresponds to the contraction of both fundamental representations and of both anti-fundamental, and $\epsilon^{a,b}$ is the totally anti-symmetric tensor.
However, the decomposition into irreducible representations of the full tensor product for this type of operator  yields 
\begin{align}
	\left( 2 \otimes \bar{2} \right) \otimes \left( 2 \otimes \bar{2} \right) = 1 \oplus 1 \oplus 3 \oplus 3 \oplus 3 \oplus 5
\end{align}
and only contains two singlets. We can associate the two singlets to the combinations $(i)$ and $(ii)$ in Eq.~(\ref{eq: U(2) quadrilinear contractions})
and we deduce that all other choices, such as~$(iii)$, must be linear combinations of these two. 
The same holds also in the $N=3$ case, where 
\begin{align}
	\left( 3 \otimes \bar{3} \right) \otimes \left( 3 \otimes \bar{3} \right) = 1 \oplus 1 \oplus 8 \oplus 8 \oplus 8 \oplus 8 \oplus 10 \oplus \overline{10} \oplus 27.
\end{align}
and the we two singlets in can be identified  with the contractions of the type $(i)$ and $(ii)$ in Eq.~(\ref{eq: U(2) quadrilinear contractions}).

Operators with the insertion of spurions proceed in a similar manner. In the $SU(2)$ case a more complicated tensor decomposition occurs 
at order~$\mathcal{O}(V^2)$ for operators with four identical left-handed fields, such as  $\left( \bar{L} L \right) \left( \bar{L} L \right)$. 
Here we find \begin{align}
	\left( 2 \otimes \bar{2} \right)^3 = 
	1 \oplus 1 \oplus 1 \oplus 1 \oplus 1 \oplus 
	3 \oplus 3 \oplus 3 \oplus 3 \oplus 3 \oplus 3 \oplus 3 \oplus 3 \oplus 3 \oplus 
	5 \oplus 5 \oplus 5 \oplus 5 \oplus 5 \oplus 
	7
\end{align}
hence we need to identify, a priopri,  five independent singlet contractions. Taking into account that 
\begin{align}
	\left( V_\ell V_\ell^\dagger \right) \otimes \left( \bar{L} L \right) \otimes \left( \bar{L} L \right) \sim 
	\left( 1 \oplus 3 \right)_{V} \otimes \left( 1 \oplus 3 \right)_{L_1} \otimes \left( 1 \oplus 3 \right)_{L_2}.
\end{align}
leads to identify the five contractions as follows
\begin{align}
	(i):\quad 1_V \otimes 1_{L_1} \otimes 1_{L_2} &\sim  V_\ell^{r} V_\ell^{\dagger \, r}  \left( \bar{L}^{s} L^{s} \right) \left( \bar{L}^{t} L^{t} \right), \nonumber \\
	(ii):\quad 1_V \otimes 3_{L_1} \otimes 3_{L_2} &\sim  V_\ell^{r} V_\ell^{\dagger \, r} \left( \bar{L}^{s} L^{t} \right) \left( \bar{L}^{t} L^{s} \right), \nonumber \\
	(iii):\quad 3_V \otimes 1_{L_1} \otimes 3_{L_2} &\sim V_\ell^{r} V_\ell^{\dagger \, s} \left( \bar{L}^{t} L^{t} \right) \left( \bar{L}^{r} L^{s} \right), \nonumber \\
	(iv):\quad 3_V \otimes 3_{L_1} \otimes 1_{L_2} &\sim V_\ell^{r} V_\ell^{\dagger \, s} \left( \bar{L}^{r} L^{s} \right) \left( \bar{L}^{t} L^{t} \right), \nonumber \\ 
	(v):\quad 3_V \otimes 3_{L_1} \otimes 3_{L_2} &\sim V_\ell^{r} V_\ell^{\dagger \, s} \left( \bar{L}^{r} L^{t} \right) \left( \bar{L}^{t} L^{s} \right).
\end{align}
The first two terms are proportional to leading order operators, while $(iii)$ and $(iv)$ are equivalent. Hence,
also in this case we have only two independent contractions that we can identify with the terms $(iii)$ and $(v)$.

\newpage

\section{Summary Tables}
\label{app:tables}

\medskip

\newcommand{\OpScale}{.85} 
\begin{table}[h]
	\centering
	\renewcommand{\arraystretch}{1.5}
	\scalebox{\OpScale}{
	\centering
	\begin{tabular}{| lc || lc | lc | lc |}
		\multicolumn{8}{c}{5--7: Fermion Bilinears} \\[.1cm] \hline
		\multicolumn{8}{|c|}{non-hermitian $(\bar L R)$} \\ \hline
		\multicolumn{2}{|c||}{5: $\psi^2 H^3 +$ h.c.} & \multicolumn{6}{c|}{6: $\psi^2 X H +$ h.c.}  \\ \hline
		$Q_{eH}$ & $(H^\dagger H)(\bar\ell_p e_r H)$ &	
		$Q_{eW}$ & $(\bar\ell_p \sigma^{\mu\nu}e_r)\tau^I H W_{\mu\nu}^I$ &
		$Q_{uG}$ & $(\bar q_p \sigma^{\mu\nu}T^A u_r)\tilde{H}G_{\mu\nu}^A$ &
		$Q_{dG}$ & $(\bar q_p \sigma^{\mu\nu}T^A d_r)H G_{\mu\nu}^A$ \\ 
		$Q_{uH}$ & $(H^\dagger H)(\bar q_p u_r \tilde{H})$ &
		$Q_{eB}$ & $(\bar\ell_p \sigma^{\mu\nu}e_r) H B_{\mu\nu}$ &
		$Q_{uW}$ & $(\bar q_p \sigma^{\mu\nu}u_r)\tau^I \tilde{H}W_{\mu\nu}^I$ &
		$Q_{dW}$ & $(\bar q_p \sigma^{\mu\nu}d_r)\tau^I H W_{\mu\nu}^I$ \\ 
		$Q_{dH}$ & $(H^\dagger H)(\bar q_p d_r H)$ &		
		& & 
		$Q_{uB}$ & $(\bar q_p \sigma^{\mu\nu}u_r)\tilde{H}B_{\mu\nu}$ &
		$Q_{dB}$ & $(\bar q_p \sigma^{\mu\nu}d_r) H B_{\mu\nu}$ \\ \hline
		\end{tabular}
	}
	\newline\centering
	\scalebox{\OpScale}{
	\begin{tabular}{| lc | lc | lc |}
		\hline \multicolumn{6}{|c|}{hermitian (+ $Q_{Hud}$) $\quad \sim \quad$ 7: $\psi^2 H^2 D$} \\ \hline
		\multicolumn{2}{|c|}{$(\bar L L)$} & 
		\multicolumn{2}{c|}{$(\bar R R)$} & 
		\multicolumn{2}{c|}{$(\bar R R^\prime)$} \\ \hline
		$Q_{H\ell}^{(1)}$ & $(H^\dagger i \overleftrightarrow{D}_\mu H)(\bar\ell_p \gamma^\mu \ell_r)$ & 
		$Q_{H e}$ & $(H^\dagger i \overleftrightarrow{D}_\mu H)(\bar e_p \gamma^\mu e_r)$ & 
		$Q_{Hud}$ + h.c. & $i(\tilde{H}^\dagger D_\mu H)(\bar u_p \gamma^\mu d_r)$ \\
		$Q_{H\ell}^{(3)}$ & $(H^\dagger i \overleftrightarrow{D}_\mu^I H)(\bar\ell_p \tau^I\gamma^\mu \ell_r)$ & 
		$Q_{H u}$ & $(H^\dagger i \overleftrightarrow{D}_\mu H)(\bar u_p \gamma^\mu u_r)$ & 
		& \\
		$Q_{Hq}^{(1)}$ & $(H^\dagger i \overleftrightarrow{D}_\mu H)(\bar q_p \gamma^\mu q_r)$ & 
		$Q_{H d}$ & $(H^\dagger i \overleftrightarrow{D}_\mu H)(\bar d_p \gamma^\mu d_r)$ & 
		& \\
		$Q_{Hq}^{(3)}$ & $(H^\dagger i \overleftrightarrow{D}_\mu^I H)(\bar q_p \tau^I\gamma^\mu q_r)$ & 
		& & 
		& \\ \hline
	\end{tabular}
	}
	\vspace{0.5cm}
	\\ \centering
	\scalebox{\OpScale}{
	\begin{tabular}{| llc | llc | llc |}
		\multicolumn{9}{c}{8: Fermion Quadrilinears} \\[.1cm] \hline
		\multicolumn{9}{|c|}{hermitian} \\ \hline
		\multicolumn{3}{|c|}{$(\bar L L)(\bar L L)$} & 
		\multicolumn{3}{c|}{$(\bar R R)(\bar R R)$} & 
		\multicolumn{3}{c|}{$(\bar L L)(\bar R R)$} \\ \hline
		$Q_{\ell\ell}$ & [a] & $(\bar\ell_p \gamma_\mu \ell_r)(\bar\ell_s \gamma^\mu \ell_t)$ & 
		$Q_{ee}$ & [a$_2$] & $(\bar e_p \gamma_\mu e_r)(\bar e_s \gamma^\mu e_t)$ & 
		$Q_{\ell e}$ & [a] & $(\bar\ell_p \gamma_\mu \ell_r)(\bar e_s \gamma^\mu e_t)$ \\
		$Q_{qq}^{(1)}$ & [a] & $(\bar q_p \gamma_\mu q_r)(\bar q_s \gamma^\mu q_t)$ & 
		$Q_{uu}$ & [a$_1$] & $(\bar u_p \gamma_\mu u_r)(\bar u_s \gamma^\mu u_t)$ & 
		$Q_{\ell u}$ & [b] & $(\bar\ell_p \gamma_\mu \ell_r)(\bar u_s \gamma^\mu u_t)$ \\
		$Q_{qq}^{(3)}$ & [a] & $(\bar q_p \gamma_\mu \tau^I q_r)(\bar q_s \gamma^\mu \tau^I q_t)$ & 
		$Q_{dd}$ & [a$_1$] & $(\bar d_p \gamma_\mu d_r)(\bar d_s \gamma^\mu d_t)$ & 
		$Q_{\ell d}$ & [b] & $(\bar\ell_p \gamma_\mu \ell_r)(\bar d_s \gamma^\mu d_t)$ \\
		$Q_{\ell q}^{(1)}$ & [b] & $(\bar\ell_p \gamma_\mu \ell_r)(\bar q_s \gamma^\mu q_t)$ & 
		$Q_{eu}$ & [b] & $(\bar e_p \gamma_\mu e_r)(\bar u_s \gamma^\mu u_t)$ & 
		$Q_{q e}$ & [b] & $(\bar q_p \gamma_\mu q_r)(\bar e_s \gamma^\mu e_t)$ \\
		$Q_{\ell q}^{(3)}$ & [b] & $(\bar\ell_p \gamma_\mu \tau^I \ell_r)(\bar q_s \gamma^\mu \tau^I q_t)$ & 
		$Q_{ed}$ & [b] & $(\bar e_p \gamma_\mu e_r)(\bar d_s \gamma^\mu d_t)$ & 
		$Q_{qu}^{(1)}$ & [a] & $(\bar q_p \gamma_\mu q_r)(\bar u_s \gamma^\mu u_t)$ \\
		& & & 
		$Q_{ud}^{(1)}$ & [b] & $(\bar u_p \gamma_\mu u_r)(\bar d_s \gamma^\mu d_t)$ & 
		$Q_{qu}^{(8)}$ & [a] & $(\bar q_p \gamma_\mu T^A q_r)(\bar u_s \gamma^\mu T^A u_t)$ \\
		& & & 
		$Q_{ud}^{(8)}$ & [b] & $(\bar u_p \gamma_\mu T^A u_r)(\bar d_s \gamma^\mu T^A d_t)$ & 
		$Q_{qd}^{(1)}$ & [a] & $(\bar q_p \gamma_\mu q_r)(\bar d_s \gamma^\mu d_t)$ \\
		& & & 
		&  & & 
		$Q_{qd}^{(8)}$ & [a] & $(\bar q_p \gamma_\mu T^A q_r)(\bar d_s \gamma^\mu T^A d_t)$ \\ \hline
	\end{tabular}
	}
	\newline\centering
	\scalebox{\OpScale}{
	\begin{tabular}{| llc | llc |}
		\hline \multicolumn{6}{|c|}{non-hermitian} \\ \hline
		\multicolumn{3}{|c|}{$(\bar L R)(\bar R L)$ + h.c.} & 
		\multicolumn{3}{c|}{$(\bar L R)(\bar L R)$ + h.c.} \\ \hline
		$Q_{\ell e d q }$ & [a] & $(\bar \ell _p^j e_r)(\bar d_s q_{tj})$ & 
		$Q_{quqd}^{(1)}$ & [b] & $(\bar q_p^j u_r)\epsilon_{jk}(\bar q_s^k d_t)$ \\ 
		& & & 
		$Q_{quqd}^{(8)}$ & [b] & $(\bar q_p^j T^A u_r)\epsilon_{jk}(\bar q_s^k T^A d_t)$ \\ 
		& & & 
		$Q_{\ell equ}^{(1)}$ & [a] & $(\bar \ell_p^j e_r)\epsilon_{jk}(\bar q_s^k u_t)$ \\ 
		& & & 
		$Q_{\ell equ}^{(3)}$ & [a] & $(\bar \ell_p^j \sigma_{\mu\nu} e_r)\epsilon_{jk}(\bar q_s^k \sigma^{\mu\nu} u_t)$ \\ \hline 
	\end{tabular}
	}
	\caption{List of all fermionic SMEFT operators in the Warsaw basis~\cite{Grzadkowski:2010es}. The division in classes is adopted from \cite{Alonso:2013hga}. 
	The letter in square brackets for the four-fermion operators labels the type of the operators as defined in section~\ref{sec:U2}.
		\label{tab:OperatorClasses}
	}
	\vglue -2 true cm
\end{table}

\begin{table}[p]
	\centering
	\renewcommand{\arraystretch}{1.2} 
	\begin{tabular}{ l c|| cc | cc | cc | cc | cc }
		\multicolumn{12}{c}{MFV breaking terms} \\[.3cm]
		Class &  No. &
		\multicolumn{2}{c|}{$Y_i^0$}
		& \multicolumn{2}{c|}{$Y_i^1$}
		& \multicolumn{2}{c|}{$Y_u^2$}
		& \multicolumn{2}{c|}{$Y_u^1 Y_d^1$}
		& \multicolumn{2}{c|}{$Y_u^2 Y_d^1$}
		\\ \hline
		$X^3,H^6,H^4 D^2,X^2 H^2$		& 15	& 9 & 6 & -- & -- & -- & -- & -- & -- & -- & --  \\ \hline\hline
		$\psi^2 H^3$								  & 3		& -- & -- & 3 & 3 & -- & -- & -- & -- & 1 & 1  \\ \hline
		$\psi^2 X H$								  & 8		& -- & -- & 8 & 8 & -- & -- & -- & -- & 3 & 3  \\ \hline
		$\psi^2 H^2 D$ 								& 8		& 7 & -- & -- & -- & 3 & -- & 1 & 1 & -- & --   \\ \hline
		total (bilinear) 								& 19 	& 7 & -- & 11 & 11 & 3 & -- & 1 & 1 & 4 & 4 \\ \hline\hline
		$(\bar{L}L)(\bar{L}L)$					& 5		& 8 & -- & -- & -- & 6 & -- & -- & -- & -- & -- \\ \hline
		$(\bar{R}R)(\bar{R}R)$				& 7		& 9 & -- & -- & -- & 5 & -- & -- & -- & -- & -- \\ \hline
		$(\bar{L}L)(\bar{R}R)$					& 8		& 8 & -- & -- & -- & 10 & -- & -- & -- & -- & --   \\ \hline
		$(\bar{L}R)(\bar{R}L)$  				& 1		& -- & -- & -- & -- & -- & -- & -- & -- & -- & --   \\ \hline
		$(\bar{L}R)(\bar{L}R)$					& 4		& -- & -- & -- & -- & -- & -- & 4 & 4 & -- & --   \\ \hline
		total (quadrilinear):						& 25 	& 25 & -- & -- & -- & 21 & -- & 4 & 4 & -- & --  \\ \\[-0.4cm] \hline\hline \\[-0.4cm]
		{\bf total:}										& 59	& 41 & 6 & 11 & 11 & 24 & -- & 5 & 5 & 4 & 4 
	\end{tabular}
	\caption{Number of independent operators in the SMEFT for the MFV scenario with insertions of different combinations of spurion.
		\label{tab:MFVdetails}
	}
\end{table}

\begin{table}[p]
	\centering
	\renewcommand{\arraystretch}{1.2} 
	\begin{tabular}{ l l c|| cc | cc | cc | cc | cc | cc}
		\multicolumn{15}{c}{ $U(2)^5$ breaking terms} \\[.3cm]
		Class & Type & No. &
		\multicolumn{2}{c|}{$V^0$}
		& \multicolumn{2}{c|}{$V^1$}
		& \multicolumn{2}{c|}{$V^2$}
		& \multicolumn{2}{c|}{$\Delta^1$}
		& \multicolumn{2}{c|}{$\Delta^1 V^1$}
		& \multicolumn{2}{c}{$V^3$}
		\\ \hline
%
		\multicolumn{2}{l}{$X^3,H^6,H^4 D^2,X^2 H^2$} & 15	& 9 & 6 & -- & -- & -- & -- & -- & -- & -- & -- & -- & -- \\ \hline\hline
		$\psi^2 H^3$ & 								& 3		& 3 & 3 & 3 & 3 & -- & -- & 3 & 3 & 3 & 3 & -- & -- \\ \hline
		$\psi^2 X H$ & 								& 8		& 8 & 8 & 8 & 8 & -- & -- & 8 & 8 & 8 & 8 & -- & -- \\ \hline
		\multirow{4}{*}{$\psi^2 H^2 D$}
			& $(\bar{L}L)$							& 4		& 8 & -- & 4 & 4 & 4 & -- & -- & -- & -- & -- & -- & -- \\
			& $(\bar{R}R)$							& 3		& 6 & -- & -- & -- & -- & -- & -- & -- & 3 & 3 & -- & -- \\
			& $Q_{Hud}$ 								& 1		& 1 & 1 & -- & -- & -- & -- & -- & -- & 2 & 2 & -- & -- \\
			& total: 								& 8		& 15 & 1 & 4 & 4 & 4 & -- & -- & -- & 5 & 5 & -- & --  \\ \hline
		\multicolumn{2}{l}{total (bilinear):}		& 19	& 26 & 12 & 15 & 15 & 4 & -- & 11 & 11 & 16 & 16 & -- & -- \\ \hline\hline
		\multirow{3}{*}{$(\bar{L}L)(\bar{L}L)$}
			& a:									& 3		& 15 & -- & 9 & 9 & 15 & 3 & -- & -- & -- & -- & 3 & 3 \\
			& b:									& 2		& 8 & -- & 8 & 8 & 12 & 4 & -- & -- & -- & -- & 4 & 4 \\
			& total:								& 5		& 23 & -- & 17 & 17 & 27 & 7 & -- & -- & -- & -- & 7 & 7 \\ \hline
		\multirow{4}{*}{$(\bar{R}R)(\bar{R}R)$}
			& a$_1$:								& 2		& 10 & -- & -- & -- & -- & -- & -- & -- & 6 & 6 & -- & -- \\
			& a$_2$:								& 1		& 3 & -- & -- & -- & -- & -- & -- & -- & 2 & 2 & -- & -- \\
			& b:									& 4 	& 16 & -- & -- & -- & -- & -- & -- & -- & 16 & 16 & -- & -- \\
			& total:								& 7 	& 29 & -- & -- & -- & -- & -- & -- & -- & 24 & 24 & -- & -- \\ \hline
		\multirow{3}{*}{$(\bar{L}L)(\bar{R}R)$}
			& a:									& 5		& 20 & -- & 10 & 10 & 10 & -- & 5 & 5 & 15 & 15 & -- & --  \\
			& b:									& 3		& 12 & -- & 6 & 6 & 6 & -- & -- & -- & 6 & 6 & -- & --  \\
			& total:								& 8		& 32 & -- & 16 & 16 & 16 & -- & 5 & 5 & 21 & 21 & -- & --  \\ \hline
		$(\bar{L}R)(\bar{R}L)$  & 					& 1		& 1 & 1 & 2 & 2 & 1 & 1 & 2 & 2 & 4 & 4 & -- & --  \\ \hline
		\multirow{3}{*}{$(\bar{L}R)(\bar{L}R)$}
			& a:									& 2		& 2 & 2 & 4 & 4 & 2 & 2 & 8 & 8 & 12 & 12 & -- & --  \\
			& b:									& 2		& 2 & 2 & 4 & 4 & 2 & 2 & 4 & 4 & 8 & 8 & -- & -- \\
			& total:								& 4		& 4 & 4 & 8 & 8 & 4 & 4 & 12 & 12 & 20 & 20 & -- & --  \\ \hline
		\multicolumn{2}{l}{total (quadrilinear):}	& 25 	& 89 & 5 & 43 & 43 & 48 & 12 & 19 & 19 & 69 & 69 & 7 & 7 \\ \\[-0.4cm] \hline\hline \\[-0.4cm]
		\multicolumn{2}{c}{\bf total:}				& 59	& 124 & 23 & 58 & 58 & 52 & 12 & 30 & 30 & 85 & 85 & 7 & 7
	\end{tabular}
	\caption{Number of independent operators in the SMEFT with $U(2)^5$ flavour symmetry and insertions of different combinations of spurion.
	\label{tab:U2detailed}}
\end{table}

\begin{table}[p]
	\centering
	\resizebox{\textwidth}{!}{
	\renewcommand{\arraystretch}{1.2} 
		\begin{tabular}{l l c||    cc | cc | cc ||    cc | cc | cc ||    cc | cc | cc ||   cc | cc ||    cc | cc ||   cc }
			\multicolumn{31}{c}{ $U(2)^5 \otimes U(1)_b \otimes U(1)_\tau$ breaking terms} \\[.3cm]
			Class & Type & No.
			& \multicolumn{2}{c|}{$V^0 $}			& \multicolumn{2}{c|}{$V^0 X^1$} 			& \multicolumn{2}{c||}{$V^0 X^2$}
			& \multicolumn{2}{c|}{$V^1$} 			& \multicolumn{2}{c|}{$V^1X^1$} 			& \multicolumn{2}{c||}{$V^1 X^2$}
			& \multicolumn{2}{c|}{$V^2$}  			& \multicolumn{2}{c|}{$V^2 X^1$} 			& \multicolumn{2}{c||}{$V^2 X^2$}
			& \multicolumn{2}{c|}{$\Delta^1$} 		& \multicolumn{2}{c||}{$\Delta^1 X^1$}
			& \multicolumn{2}{c|}{$\Delta^1 V^1$} 	& \multicolumn{2}{c||}{$\Delta^1 V^1 X^1 $}
			& \multicolumn{2}{c}{$ V^3$}
			\\ \hline
			\multicolumn{2}{l}{$X^3,H^6,H^4 D^2,X^2 H^2$} & 15
			& 9 & 6 	& -- & -- 	& -- & --
			& -- & --	& -- & -- 	& -- & --
			& -- & --	& -- & -- 	& -- & --
			& -- & --	& -- & --
			& -- & --	& -- & --
			& -- & --	\\ \hline\hline
			$\psi^2 H^3$ & & 3
			& 1	& 1 	& 2 & 2		& -- & --
			& 1	& 1 	& 2 & 2  	& -- & --
			& -- & --	& -- & -- 	& -- & --
			& 3	& 3		& -- & --
			& 3	& 3		& -- & --
			& -- & -- \\ \hline
			$\psi^2 X H$ & & 8
			& 3 & 3 	& 5 & 5  	& -- & --
			& 3	& 3		& 5 & 5 	& -- & --
			& -- & -- 	& -- & -- 	& -- & --
			& 8 & 8 	& -- & --
			& 8 & 8 	& -- & --
			& -- & -- \\ \hline
			\multirow{4}{*}{$\psi^2 H^2 D$}
			& $(\bar{L}L)$ & 4
			& 8 & -- 	& -- & --	& -- & --
			& 4	& 4 	& -- & --	& -- & --
			& 4 & -- 	& -- & -- 	& -- & --
			& -- & -- 	& -- & --
			& -- & --	& -- & --
			& -- & --  \\
			& $(\bar{R}R)$ & 3
			& 6 & -- 	& -- & --	& -- & --
			& -- & -- 	& -- & --	& -- & --
			& -- & -- 	& -- & -- 	& -- & --
			& -- & -- 	& -- & --
			& 1 & 1		& 2 & 2
			& -- & -- \\
			& $Q_{Hud}$ & 1
			& -- & -- 	& 1 & 1 	& -- & --
			& --& -- 	& -- & --	& -- & --
			& -- & -- 	& -- & -- 	& -- & --
			& -- & -- 	& -- & --
			& 1 & 1		& 1 & 1
			& -- & -- \\
			& total: & 8
			& 14 & -- 	& 1 & 1 	& -- & --
			& 4	& 4 	& -- & --	& -- & --
			& 4 & -- 	& -- & -- 	& -- & --
			& -- & -- 	& -- & --
			& 2 & 2		& 3 & 3
			& -- & -- \\ \hline
			\multicolumn{2}{l}{total (bilinear):} & 19
			& 18 & 4 	& 8 & 8 	& -- & --
			& 8	& 8 	& 7 & 7		& -- & --
			& 4 & -- 	& -- & -- 	& -- & --
			& 11 & 11 	& -- & --
			& 13 & 13	& 3 & 3
			& -- & --
			\\\hline\hline
			\multirow{3}{*}{$(\bar{L}L)(\bar{L}L)$}
			& a:
			& 3
			& 15 & -- 	& -- & -- 	& -- & --
			& 9	& 9 	& -- & -- 	& -- & --
			& 15 & 3	& -- & -- 	& -- & --
			& -- & -- 	& -- & --
			& -- & -- 	& -- & --
			& 3 & 3 \\
			& b:		& 2
			& 8 & -- 	& -- & -- 	& -- & --
			& 8 & 8  	& -- & -- 	& -- & --
			& 12 & 4 	& -- & -- 	& -- & --
			& -- & -- 	& -- & --
			& -- & -- 	& -- & --
			& 4 & 4 \\
			& tot.: 	& 5
			& 23 & -- 	& -- & -- 	& -- & --
			& 17 & 17 	& -- & -- 	& -- & --
			& 27 & 7 	& -- & -- 	& -- & --
			& -- & -- 	& -- & --
			& -- & -- 	& -- & --
			& 7 & 7  \\\hline
			\multirow{4}{*}{$(\bar{R}R)(\bar{R}R)$}
			& a$_1$:		& 3
			& 10 & -- 	& -- & -- 	& -- & --
			& -- & -- 	& -- & -- 	& -- & --
			& -- & -- 	& -- & -- 	& -- & --
			& -- & -- 	& -- & --
			& 3 & 3 	& 3 & 3
			& -- & -- \\
			& a$_2$:		& 3
			& 3 & -- 	& -- & -- 	& -- & --
			& -- & -- 	& -- & -- 	& -- & --
			& -- & -- 	& -- & -- 	& -- & --
			& -- & -- 	& -- & --
			& -- & -- 	& 2 & 2
			& -- & -- \\
			& b:		& 4
			& 16 & -- 	& -- & -- 	& -- & --
			& -- & -- 	& -- & -- 	& -- & --
			& -- & -- 	& -- & -- 	& -- & --
			& -- & -- 	& -- & --
			& 6 & 6 	& 10 & 10
			& -- & -- \\
			& tot.:	& 7
			& 29 & -- 	& -- & -- 	& -- & --
			& -- & -- 	& -- & -- 	& -- & --
			& -- & -- 	& -- & -- 	& -- & --
			& -- & -- 	& -- & --
			& 9 & 9 	& 15 & 15
			& -- & -- \\\hline
			\multirow{3}{*}{$(\bar{L}L)(\bar{R}R)$}
			& a:		& 5
			& 20 & -- 	& -- & -- 	& -- & --
			& 10 & 10 	& -- & -- 	& -- & --
			& 10 & -- 	& -- & -- 	& -- & --
			& 2 & 2 	& 3 & 3
			& 6 & 6 	& 9 & 9
			& -- & -- \\
			& b:		& 3
			& 12 & -- 	& -- & -- 	& -- & --
			& 6 & 6 	& -- & -- 	& -- & --
			& 6 & -- 	& -- & -- 	& -- & --
			& -- & -- 	& -- & --
			& 2 & 2 	& 4 & 4
			& -- & -- \\
			& tot.:	& 8
			& 32 & -- 	& -- & -- 	& -- & --
			& 16 & 16 	& -- & -- 	& -- & --
			& 16 & -- 	& -- & -- 	& -- & --
			& 2 & 2 	& 3 & 3
			& 8 & 8 	& 13 & 13
			& -- & -- \\\hline
			$(\bar{L}R)(\bar{R}L)+h.c.$
			& tot.:	& 1
			& --  & --  & -- & -- 	& 1 & 1
			& -- & -- 	& -- & -- 	& 2 & 2
			& -- & -- 	& -- & -- 	& 1 & 1
			& -- & -- 	& 2 & 2
			& -- & -- 	& 4 & 4
			& -- & --  \\\hline
			\multirow{3}{*}{$(\bar{L}R)(\bar{L}R)+h.c.$}
			& a:		& 2
			& -- & -- 	& 2 & 2 	& -- & --
			& -- & -- 	& 4 & 4 	& -- & --
			& -- & -- 	& 2 & 2 	& -- & --
			& 2 & 2 	& 2 & 2
			& 4 & 4 	& 4 & 4
			& -- & --  \\
			& b:		& 2
			& -- & -- 	& 2 & 2 	& -- & --
			& -- & -- 	& 4 & 4 	& -- & --
			& -- & -- 	& 2 & 2 	& -- & --
			& 4 & 4 	& 4 & 4
			& 6 & 6 	& 6 & 6
			& -- & --  \\
			& tot.:	& 4
			& -- & -- 	& 4 & 4 	& -- & --
			& -- & -- 	& 8 & 8 	& -- & --
			& -- & -- 	& 4 & 4 	& -- & --
			& 6 & 6 	& 6 & 6
			& 10 & 10 	& 10 & 10
			& -- & --
			\\ \hline
			\multicolumn{2}{l}{total (quadrilinear):}	& 25
			& 84 & --	& 4 & 4 	& 1 & 1
			& 33 & 33	& 8 & 8 	& 2 & 2
			& 43 & 7	& 4 & 4		& 1 & 1
			& 8 & 8		& 11 & 11
			& 27 & 27 	& 42 & 42
			& 7 & 7 \\ \\[-0.4cm] \hline\hline \\[-0.4cm]
			\multicolumn{2}{c}{\bf total:} & 59
			& 111 & 10 	& 12 & 12 	& 1 & 1
			& 41 & 41 	& 15 & 15 	& 2 & 2
			& 47 & 7 	& 4 & 4		& 1 & 1
			& 19 & 19 	& 11 & 11
			& 40 & 40 	& 45 & 45
			& 7 & 7
		\end{tabular}
	}
	\caption{Number of independent operators in the SMEFT with $U(2)^5 \otimes U(1)^2$ flavour symmetry and insertions of different combinations of spurion.
	\label{tab:U2U1detailed}}
\end{table}

\begin{table}[h]
	\centering
	\renewcommand{\arraystretch}{1.2}
	\begin{tabular}{*{10}{p{1.2cm}}}
		& (11) & (12) & (13) & (21) & (22) & (23) & (31) & (32) & (33)
	\end{tabular}\newline\centering
	\begin{tabular}{l|*{9}{p{1.2cm}|}}
		\cline{2-10}
		(11) 	&	$a_1\quad$ $a_2$ 	&  &  &  & $2 a_1$ $c_1 \epsl^2$ & $ \beta_1 \epsl$ &  & $\beta_1^\ast  \epsl$ & $a_3$ \\[1cm] \cline{2-10}
		(12) 	&	 &  &  & $2 a_2$ $c_4  \epsl^2$ &  &  & $\beta_3^\ast \epsl$ &  &  \\[1cm] \cline{2-10}
		(13)	&	 &  &  & $\beta_3 \epsl$ &  &  & $a_4$ &  &  \\[1cm] \cline{2-10}
		(21)	&	 & $2 a_2$ $c_4  \epsl^2$ & $\beta_3 \epsl $ &  &  &  &  &  &  \\[1cm] \cline{2-10}
		(22)	&	 $2 a_1$ $c_1 \epsl^2$ &  &  &  & $a_1\quad$  $a_2$ $c_1  \epsl^2$ $c_4  \epsl^2$ & $\beta_1  \epsl$ $\beta_3  \epsl $ $\xi_1  \epsl^3 $ &  & $\beta_1^\ast 
		\epsl$ $\beta_3^\ast \epsl$ $\xi_1^\ast \epsl^3$ & $a_3$ \newline  $c_2  \epsl^2$ \\[1cm] \cline{2-10}
		(23)	&	$\beta_1 \epsl$ &  &  &  & $\beta_1  \epsl $ $\beta_3 \epsl$ $\xi_1 \epsl^3$  & $\gamma_1 \epsl^2$ &  & $a_4$ \newline $c_3  \epsl^2$ & $\beta_2 \epsl$ \\[1cm] \cline{2-10}
		(31)	&	 & $\beta_3^\ast\epsl $ & $a_4$ &  &  &  &  &  &  \\[1cm] \cline{2-10}
		(32)	&	$\beta_1^\ast \epsl$ &  &  &  & $\beta_1^\ast  \epsl$ $\beta_3^\ast \epsl$ $\xi_1^\ast \epsl^3$ & $a_4$ \newline $c_3  \epsl^2$ &  & $\gamma_1^\ast \epsl^2$ & $\beta_2^\ast \epsl $ \\[1cm] \cline{2-10}
		(33)	&	$a_3$ &  &  &  & $a_3$ \newline $c_2  \epsl^2$ & $\beta_2\epsl $ &  & $\beta_2^\ast \epsl$ & $a_5$ \\[1cm] \cline{2-10}
	\end{tabular}
	\caption{The $\Sigma_{\ell \ell}^{ij,nm}$ tensor in the interaction basis as defined in Eq.~(\ref{eq:SigmaTens}): the entries are as indicated in rows ($ij$) and columns ($nm$), respectively. All terms in each cell should be added.
\label{tab:SigmaLL} }
\end{table}

\begin{table}[p]
	\centering
	\renewcommand{\arraystretch}{1.2}
	\begin{tabular}{*{10}{p{1.2cm}}}
		& (11) & (12) & (13) & (21) & (22) & (23) & (31) & (32) & (33)
	\end{tabular}\newline\centering
	\begin{tabular}{l|*{9}{p{1.2cm}|}}
		\cline{2-10}$ $
		(11) 	&	 $a_1$ 	&  &  &  & $a_1 \quad$  $c_3\epsq^2$ & $\beta_3 \epsq$ &  & $\beta_3^\ast \epsq$ & $a_2$ \\[1cm] \cline{2-10}
		(12) 	&	 &  &  &  &  &  &  &  &  \\[1cm] \cline{2-10}
		(13)	&	 &  &  &  &  &  &  &  &  \\[1cm] \cline{2-10}
		(21)	&	 &  &  &  &  &  &  &  &  \\[1cm] \cline{2-10}
		(22)	&	 $a_1\quad$ $c_1 \epsl^2$ &  &  &  & $a_1\quad$ $c_1 \epsl^2$ $c_3 \epsq^2$ & $\beta_3 \epsq$ $\xi_1 \epsl^2 \epsq$ &  & $\beta_3^\ast \epsq $ $\xi_1^\ast  \epsl^2 \epsq$ & $a_2\quad$ $c_2 \epsl^2$ \\[1cm] \cline{2-10}
		(23)	&	 $\beta_1 \epsl$ &  &  &  & $\beta_1 \epsl$ $\xi_2  \epsq^2 \epsl $ & $\gamma_1 \epsl \epsq$ &  & $\gamma_2  \epsl \epsq$ & $\beta_2 \epsl$ \\[1cm] \cline{2-10}
		(31)	&	 &  &  &  &  &  &  &  &  \\[1cm] \cline{2-10}
		(32)	&	 $\beta_1^\ast \epsl$ &  &  &  & $\beta_1^\ast \epsl$ $\xi_2^\ast  \epsq^2 \epsl $ & $\gamma_2^\ast  \epsl \epsq$ &  & $\gamma_1^\ast  \epsl \epsq$ & $\beta_2^\ast \epsl$ \\[1cm] \cline{2-10}
		(33)	&	 $a_3$ &  &  &  & $a_3\quad $ $c_4 \epsq^2$ & $\beta_4 \epsq$ &  & $\beta_4^\ast \epsq$ & $a_4$ \\[1cm] \cline{2-10}
	\end{tabular}
	\caption{The $\Sigma_{\ell q}^{ij,nm}$ tensor in the interaction basis as defined in Eq.~(\ref{eq:SigmaTens}): the entries are as indicated in rows ($ij$) and columns ($nm$), respectively. All terms in each cell should be added.
	\label{tab:SigmaQL} }
\end{table}

\end{document}